\documentclass[12pt]{report}
\usepackage{latexsym}
\usepackage{graphicx}
\usepackage[thmmarks]{ntheorem}
\usepackage{algorithm} 
\usepackage{subfigure}
\usepackage{tabularx}
\usepackage{bibref}
\usepackage{calc}
\usepackage{xspace}
\usepackage{fancyhdr}
\usepackage{afterpage}
\usepackage{lscape}
\usepackage{setspace}
\normalsize 
\parskip \medskipamount
\floatstyle{plain}
\restylefloat{figure}
\floatstyle{plain}
\restylefloat{table}

\newcommand{\urlBiBTeX}[1]{\url{#1}}
{\theoremsymbol{\ensuremath{\Box}}%
\theorembodyfont{\rmfamily}%
\theoremstyle{nonumberplain}\theoremseparator{:}%
}
\begin{document}

\begin{singlespace}

%
%
%
   \newcommand{\theauthor}{Annie Thomas}
   \newcommand{\thetitle}{Three dimensional chaos game representation of protein sequences}
   \newcommand{\thedegree}{Master of Science}
   \newcommand{\exammonth}{May}  
   \newcommand{\examyear}{2005}      
   \newcommand{\finalyear}{2005}     
\pagenumbering{roman}
\thispagestyle{empty}%
\vbox{
   \begin{center}
      {\LARGE \thetitle\\}
      \vspace{1.0in}
      by\\
      \vspace{0.75in}
      {\Large \theauthor\\}
      \vspace{1.25in}
      {\large Graduate Program in Computer Science}\\
   \end{center}
}
\vbox{
   \begin{center}
      Submitted in partial fulfillment\\
      of the requirements for the degree of\\
      \thedegree\\
      \vspace{6em}
      Faculty of Graduate Studies\\
      The University of Western Ontario\\
      London, Ontario\\
      \exammonth, \finalyear\\
   \end{center}
}
\vbox{
   \begin{center}
      \copyright~\theauthor~\finalyear
   \end{center}
}
\newpage

\setcounter{page}{2}
\newcommand{\signwidth}{.45\textwidth}
\newcommand{\datewidth}{.35\textwidth}
\newcommand{\signheight}{.002in}
\newcommand{\signature}{\rule[-.25in]{\signwidth}{\signheight}}
\newcommand{\dateline}{\rule[-.25in]{\datewidth}{\signheight}}

\vfil\vbox{
   \begin{center}
      THE UNIVERSITY OF WESTERN ONTARIO\\
      FACULTY OF GRADUATE STUDIES\\[\baselineskip]

      CERTIFICATE OF EXAMINATION
   \end{center}
}\vfil%

\vbox{
   \begin{tabular}{p{\signwidth} c p{\signwidth}}
     Supervisor                 & &    Examiners\\
     \signature                 & &    \signature\\
     \vspace{.008in}Co-Supervisor              & &    \signature\\
     \signature                 & &    \signature\\
     \vspace{.008in}Supervisory Committee      & &    \signature\\
     \signature                 & &     \\

   \end{tabular}
}\vfil%

\vbox{
   \begin{center}
     The thesis by\\ \vspace{.15in}
     \theauthor\\ \vspace{.15in}
     entitled\\
      \vspace{.25in}
      \textsc{\thetitle}\\
      \vspace{.25in}
      is accepted in partial fulfillment of the\\
      requirements for the degree of\\
      \thedegree
   \end{center}
}\vfil%

\vbox{\noindent%
   \begin{tabular}{p{\signwidth} c p{\signwidth}}
   \dateline & &    \signature\\
     Date              & &   Chair of Examining Board\\
   \end{tabular}
}\vfil\vbox{}

\end{singlespace}

\chapter*{Abstract}

\paragraph
\indent A new three dimensional approach to the chaos game
representation of protein sequences is explored in this thesis.
The basics of DNA, the synthesis of proteins from DNA, protein
structure and functionality and sequence alignment techniques are
presented. The mathematical background needed for understanding
the chaos game representation and fractal analysis are briefly
discussed.
\paragraph
\indent An account of the existing literature on the chaos game
representation of DNA sequences and a detailed account of the
chaos game representation of protein sequences in two dimensions
with its advantages and limitations are presented. We explore a
new three dimensional approach to the chaos game representation of
protein sequences (3D-CGR) and study its ability \textit{a)} to
determine protein sequence similarity and differences, \textit{b)}
to study the  effect of dinucleotide biases at amino acid level on
the 3D-CGR derived protein homology, and \textit{c)} to identify
sequence similarity based on shuffled motifs that could be used
for studying protein evolution due to exon shuffling.

\emph{Keywords:  Chaos game, dinucleotide bias, protein homology,
motifs, exon shuffling, fractal}

\chapter*{Acknowledgments}
\paragraph
\indent This thesis would not have been possible without the
support of many people. I thank my supervisor Dr.Lila Kari for her
time, suggestions, patience and encouragement. I thank my
co-supervisor Dr.Kathleen Hill for her time, sharing her expert
knowledge in Molecular Biology and providing suggestions and
interpretations at every stage of the thesis.

\paragraph
\indent I also especially thank Dr.Fernando Sancho for his help
with fractal geometry concepts in chapter 3 and reading the
thesis. I am grateful to a very special person Dr. Rani
Siromoney, for her support, encouragement and prayer. Also, I
thank two very special people in my life, my aunt Anne and uncle
Daya for being there for me always.

\paragraph
\indent And finally, thanks to my parents for their unconditional
love, my husband,  brother and uncle Baga  for their suggestions, support and
encouragement, and numerous friends who endured this long process
with me, always offering support and love.


\tableofcontents

\listoftables

\listoffigures



\chapter{Introduction}
\label{chap:introduction}

%
%
\fancypagestyle{plain}{%
  \fancyhf{}%
  \fancyhead[R]{\thepage}%
  \renewcommand{\headrulewidth}{0pt}%
}
%
%
\fancypagestyle{headings}{%
  \fancyhf{}%
  \fancyhead[R]{\thepage}%
  \renewcommand{\headrulewidth}{0pt}%
}

%
%
\pagestyle{headings} 

\pagenumbering{arabic}

%
%

\paragraph
\indent  Protein sequence analysis is the key tool for
understanding the evolution of proteins, sequence classification
and for identifying conserved (they remain same across all
species) positions crucial for the function and structure of
proteins. This thesis is intended to study the protein sequence
similarity using a holistic approach differing from the
traditional sequence alignment one which is based on subsequences.
The tool used for studying the sequence as a whole is known as
\textit{Chaos Game Representation} (CGR).
\paragraph
\indent Jeffrey in 1990 \cite{Jeffrey1990} introduced the new tool
CGR to visually represent DNA sequences. This new tool stimulated
interest among researchers and CGR has since been used to explore
the primary sequence organization of DNA and proteins. Although
research on CGR has been widely explored, it has been limited to a
two dimension representation. This thesis goes a step further to
represent the CGR in three dimensions and to understand its
potential as a tool in analysing protein sequence similarity.
\paragraph
\indent Following the introduction, Chapter 2 gives the basics of
molecular biology: The basic notions of DNA and protein sequences,
the synthesis of protein from DNA and the representation of
sequence/species relatedness through phylogenetic trees. It also
looks into bioinformatic techniques such as sequence alignment and
multiple sequence alignment with CLUSTALW, needed for the
understanding of the thesis.
\paragraph
\indent Chapter 3 looks into the mathematics behind the chaos game
representation. This chapter explains what the chaos game is, how
it could reveal the underlying patterns of any
sequence. Also, this chapter gives a brief introduction into the mathematics of
generating fractals.
\paragraph
\indent In Chapter 4, the usefulness of the CGR explored in the
past is explained. The literature on the CGR is grouped into two
sections, one on DNA sequences and the other on protein sequences.
Since the emphasis of the thesis is on protein sequences, a
detailed analysis of all the previous work of the CGR of protein
has been provided.
\paragraph
\indent Chapter 5 deals with the new three dimensional approach to
the CGR and results of the analysis of protein sequences using
3D-CGR. In the beginning, the objectives of the new approach are
presented: to detect protein homology using 3D-CGR, to understand
the impact of dinucleotide bias at the amino acid level on the
3D-CGR derived protein homology and to study the sequence
relatedness to detect shuffled motifs. Following this, a
description of the geometric solid icosahedron used for playing
the chaos game, the method of representing amino acids on the
icosahedron, and the chaos game in three dimensions are explained.
Also, various distance measures used in the thesis and the spatial
subdivision method used for determining the fractal dimension are
explained in this chapter.
\paragraph
\indent Next, the experimental objectives  of the thesis are
discussed. They are: (i) the validation of the phylogenetic trees
obtained using 3D-CGR for detecting sequence relatedness, (ii)
detection of the impact of dinucleotide bias at the amino acid
level on the 3D-CGR derived protein homology, (ii) comparison of
the trees generated by the 3D-CGR and CLUSTALW for sequence
relatedness and shuffled motif detection, (iv) comparison of the
effect of using various distance measures on the phylogenetic
trees and (v) study the sequence relatedness using fractal
patterns.
\paragraph
\indent Following this, the methodologies used for performing the
experiments are presented in detail. Our experiments reveal that
the 3D-CGR can distinguish protein families and species
relatedness within the families of the sequences. The 3D-CGR can
detect shuffled motifs that cannot be detected by CLUSTALW. The
detection of shuffled motifs by 3D-CGR could be a useful tool in
studying protein evolution due to exon shuffling. Also, the
significant difference in branch length between closely related
sequences on comparison with CLUSTALW indicate that 3D-CGR could
be used for measuring the amount of divergence between sequences
within a family. The impact of dinucleotide bias at the amino acid
level was seen in the branch length between some of the closely
related sequences and in the branch order of the families.
Finally, the sequence relatedness assessed using fractal curves
and its limitation in studying protein homology is explained.
\paragraph
\indent Lastly, Chapter 6 concludes the thesis by briefly presenting
the major concepts discussed in each chapter, the novel outcome of the
new approach and few words on the future work using 3D-CGR.

\chapter{DNA and protein sequences}
\label{chap:Bioint}

\section{DNA}
\paragraph
\indent Deoxyribonucleic acid (DNA) stores the genetic information
that determines all activities of every living organism. The
genetic information stored in DNA is passed from one generation to
the next. DNA is made up of four \textit{nucleotides:} guanine,
adenine, thymine and cytosine, often referred as G,A,T,C.
Nucleotides are organic structures made up of three subunits:
phosphate, deoxyribose sugar and a nitrogenous base. The four
nucleotides G,A,T and C have the same phosphate and sugar group
but differ in their nitrogenous bases (fig 2.1).
\subsection{Structure}
\paragraph
\indent Nucleotides are linked to each  other by the phosphate
group of one nucleotide with the deoxyribose sugar of another
nucleotide forming a strand. The hydroxyl groups on the 5'(5th
carbon)- and 3'(3rd carbon) of deoxyribose sugar link to the
phosphate groups to form the DNA backbone. DNA is a double
stranded molecule with the two strands in \textit{anti-parallel}
directions. The 5' end of one strand corresponds to the 3' end of
the complementary strand and vice versa \cite{Krane2003}. The
strands are hydrogen bonded together by the base pairs A-T and
G-C, the nucleotide A in one strand is hydrogen bonded with
nucleotide T in the other strand and similarly nucleotide C in one
strand is hydrogen bonded with nucleotide G on the other. For
example, if one strand is 5'-ACTG-3' then the other strand is
3'-TGAC-5'. These strands twist together to form a double helical
structure.
\begin{figure}
\begin{center}
\includegraphics[scale=0.50]{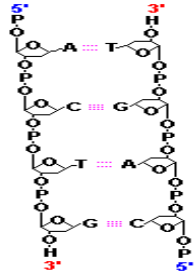}
\caption{DNA; A, T, G, C - nucleotides; Anti-parallel strands (5'
to 3' and 3' to 5') bonded by base pairs A-T and G-C;}
\label{fig:2.1}
\end{center}
\end{figure}
\subsection{ Genetic code}
\label{sect:gencode}
\paragraph
\indent A strand of DNA is composed of coding and non-coding
regions. A coding region refers to part of a DNA strand that
contains the genetic information necessary for producing the amino
acid chains of proteins responsible for performing many cellular
functions. The non-coding regions on the other
hand do not participate in amino acid chain formation. The process
of generating proteins from DNA is know as \textit{gene
expression} and this process involves two stages,
\textit{transcription} and \textit{translation} (fig 2.2).
\begin{figure}
\begin{center}
\includegraphics[scale=0.50]{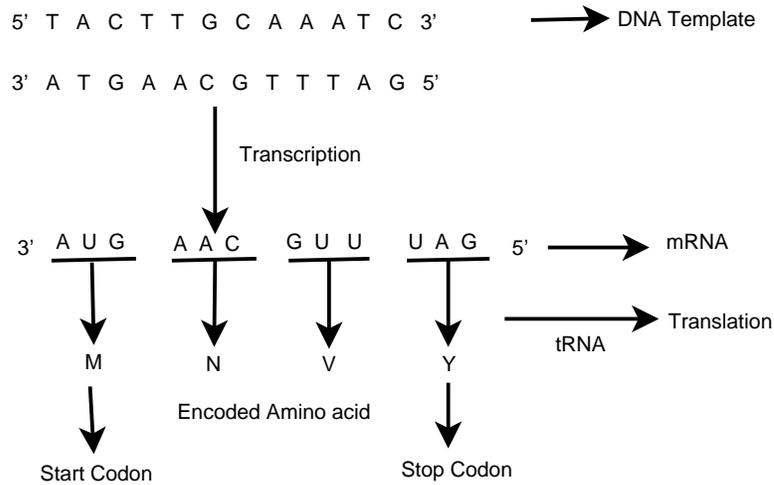}
\caption{Transcription and translation} \label{fig:2.2}
\end{center}
\end{figure}
\paragraph
\indent In \textit{transcription}, the nucleotide sequence that
ultimately encodes a protein is used as a template to code for RNA (ribonucleic
acid), known as mRNA (\textit{messenger RNA}). RNA molecules are
similar to DNA except the deoxyribose sugar in DNA is replaced by
ribose in RNA, the base Thymine (T) in DNA is replaced by the base
Uracil (U) in RNA and RNA is a single stranded structure. The
process by which mRNA codes for protein is translation. The
translation process is performed by a special type of RNA called
tRNA(\textit{transfer RNA}) that deciphers triplet nucleotide code of
mRNA to specific amino acids. These triplet nucleotides are
referred as \textit{codons}. The relationship between codon and
amino acid is referred as \textit{Genetic code} (fig 2.3). Since
DNA is composed of four nucleotides, there are {$4^{3}$=64}
possible codons. The beginning of translation is signaled by a
special codon called \textit{start codon}. There are three codons
that do not encode any amino acid but instead signal the end of
translation, they are called \textit{Stop Codons}. Since there
are only twenty amino acids that make up proteins, more than one
codon may refer to a particular amino acid.
\begin{figure}
\begin{center}
\includegraphics[scale=0.45]{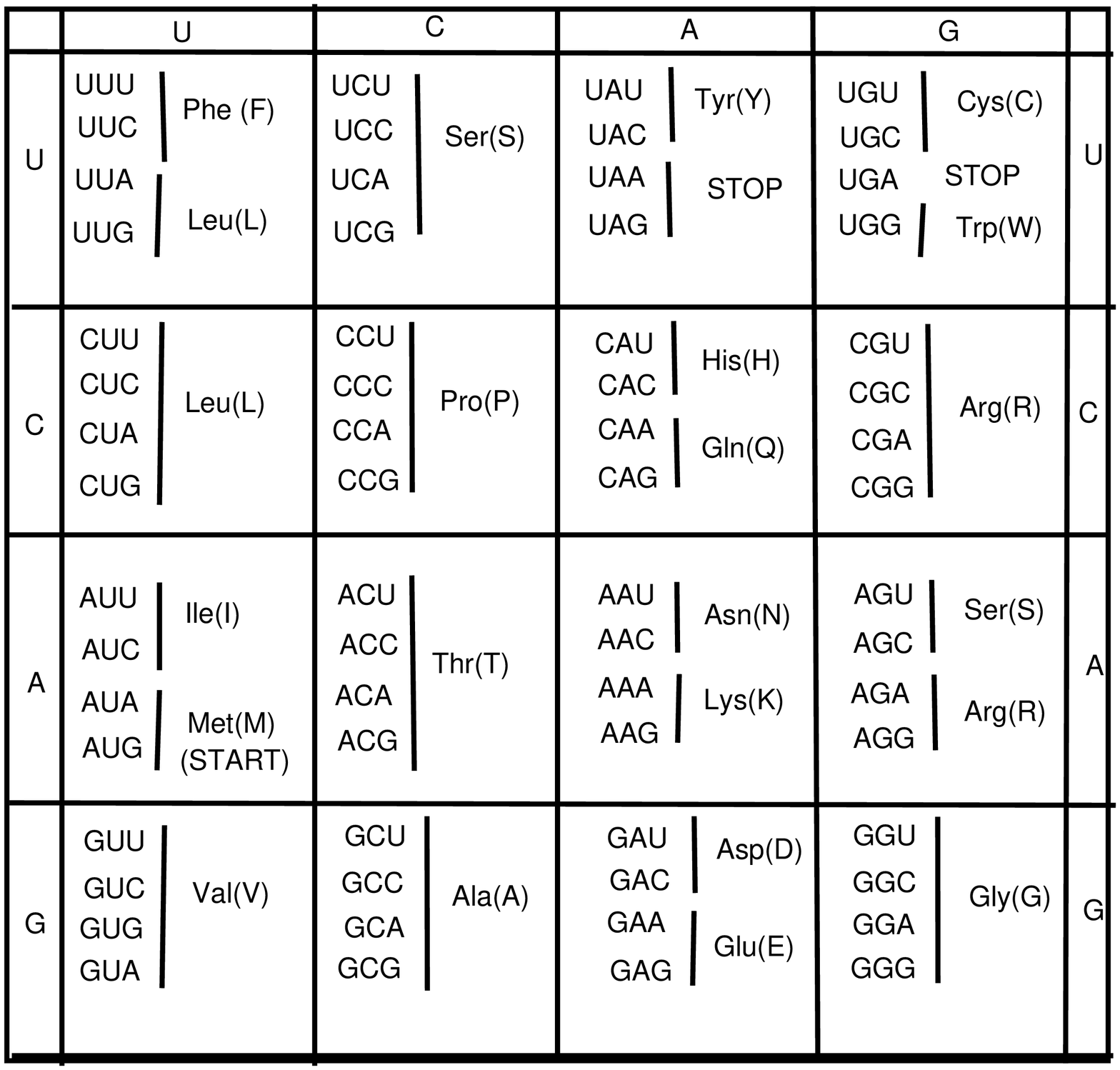}
\caption{Genetic code - Triplet codon (example: UUU); 3 letter
representation of amino acid (example: Phe) and corresponding 1
letter representation (example: F)} \label{fig:2.3}
\end{center}
\end{figure}
\section{Proteins}
\paragraph
\indent Proteins are long chain molecules built from twenty amino
acids encoded by \textit{Codons}. The twenty amino acids are
Alanine(A), Arginine(R), Asparagine(N), Aspartic acid(D),
Cysteine(C), Glutamic acid(E), Glutamine(Q), Glycine(G),
Histidine(H), Isoleucine(I), Leucine(L), Lysine(K),
Methionine(M),Phenylalanine(F), Proline(P), Serine(S),
Threonine(T), Tryptophan(W), Tyrosine(Y) and Valine(V). \paragraph
\indent Amino acids are small molecules containing an amino group,
carboxyl group, hydrogen atom and a side chain (or R group)
attached to the central carbon (or alpha carbon)
\cite{Zimmermann2003}. Amino acids differ only in the side chain
R. The amino acids are linked to one another by the central carbon
of one amino acid with the amino group of the other amino acid
forming a peptide bond. The amino acids in the long chains are
referred as \textit{residues}.
\begin{figure}
\begin{center}
\includegraphics{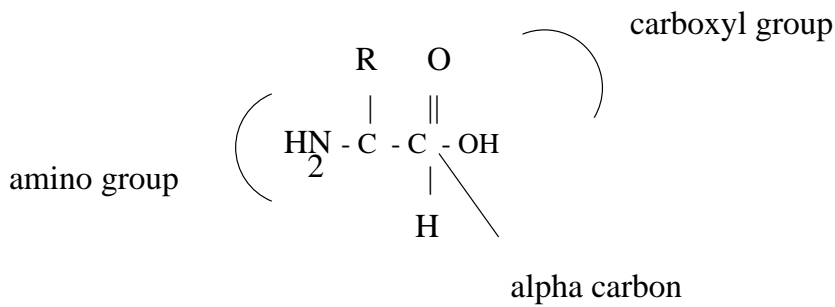}
\caption{The basic structure of an amino acid} \label{fig:2.4}
\end{center}
\end{figure}
\subsection{Amino acid classification}
\paragraph
\indent The similarity of amino acids is classified based on the
chemical properties of the side chain ( R group). Two major
classifications are Hydrophobic (non-polar) - not soluble in
water: A, I, L, M, F, P, W and V and Hydrophilic (polar) - soluble
in water: R, H, K, D, E, N, C, Q, G, S, T and Y. The polar class
is further divided into: positively charged: R, H and K,
negatively charged: D and E and uncharged: N, C, Q, G, S, T and Y.
Amino acids are also grouped together based on their
physio-chemical properties by which they could be substituted for
one another in protein sequences with minimal apparent affect on
the functionality of the proteins, known as \textit{Conservative
substitutions} \cite{Basu1997}. The following are some of the
groupings: ILVM, RK, DE, ST, AG and FY.
\subsection{Structure}
\paragraph
\indent The sequence of amino acids forming a protein is known as
the primary structure of the protein. In nature, protein molecules
collapse and fold into a unique structure known as \textit{native
structure}. There are some patterns in the native structure that
are quite common and found in many proteins, the location and
direction of these patterns are called \textit{secondary
structures}. The three main secondary structures are
$\alpha$-helix, $\beta$-sheet and random coil.  The $\alpha$-helix
is formed by the hydrogen bonds between the carbonyl group of the
$ith$ residue and the nitrogen group of the $(i+4)th$ residue (fig
2.5). An $\alpha$-helix on average has 10 residues having 3.6
residues per turn \cite{Krane2003}. $\beta$-sheets are strands of
amino acid sequences forming hydrogen bonds between them. The
bonds are formed between the carbonyl oxygen of amino acids from
one strand with nitrogen groups of the other strand. The strands
could be parallel or anti-parallel to each other. A $\beta$-sheet
can consist of all parallel strands or all anti-parallel strands
or can contain both. $\beta$-sheets usually consist of 5 to 10
residues \cite{Philip2003}(fig 2.6). Random coils are sequences of
amino acids that connect  $\alpha$- helices and $\beta$-sheets and
they are not regular structures, both in shape and size.
\begin{figure}
\begin{center}
\includegraphics[width=0.4\textwidth, height=0.4\textheight]{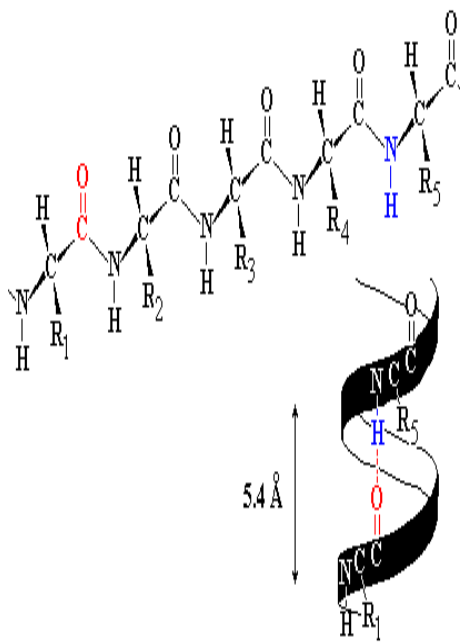}
\caption{Protein secondary structure: $\alpha$- Helix}
\label{fig:2.5}
\end{center}
\end{figure}
\subsection{Motif and domain}
\paragraph
\indent A motif is a combination of a few secondary structures
\cite{Philip2003}. For example, \textit{helix - random coil -
helix} is a motif. A domain is a more complex combination of
secondary structures having a specific function by binding to
external molecules (i.e, DNA), therefore referred to as active
site. A domain could maintain its characteristic structure even if
separated from the original protein \cite{Philip2003}. A protein
can have several motifs, which can combine to form specific
domains, and one or more domains together form the protein's
tertiary structure.
\begin{figure}
\begin{center}
\includegraphics[width=0.4\textwidth, height=0.4\textheight]{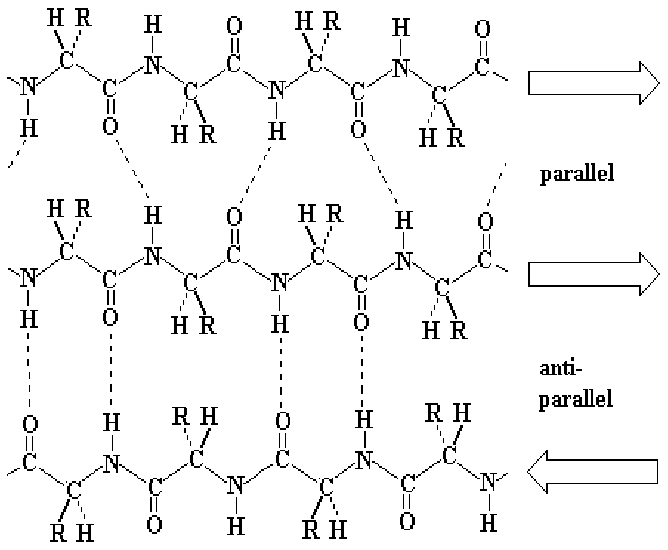}
\caption{Protein secondary structure: $\beta$- Sheets}
\label{fig:2.6}
\end{center}
\end{figure}
\subsection{Sequence alignment}
\subsubsection{Pairwise alignment}
\paragraph
\indent The number of proteins that exists in nature is very large
but these proteins could be classified  based on the sequence
pattern, structure and their functionality. Proteins that have
similarities in their sequences are believed to be derived from an
common ancestor. Therefore, determining the similarities of
sequences would help to understand the evolution of proteins.
Sequence alignment is a technique used to compare biological
sequences in order to study the similarities and differences to
find the origin of evolution between them. Sequences that share
sequence similarity would differ from one another in some sequence
positions due to
\begin{enumerate}
    \item substitution that replaces one nucleotide/amino acid with another.
    \item an insertion that adds one or more nucleotides/amino acids.
    \item deletion that deletes one or more nucleotides/amino acids.
    \item inversion that reverses the orientation of subsequences.
\end{enumerate}
\paragraph
\indent Given two sequences $a = a_{1}\ldots a_{m}$ and $b=
b_{1}\ldots b_{n}$ over the alphabet $\Sigma$. An alignment of the
sequences $a$ and $b$ is a pair of sequences $a^{'}_{1} \ldots
a^{'}_{l}$ and $b^{'}_{1} \ldots b^{'}_{l}$ of equal lengths
defined over the extended alphabet $\Sigma'$ = $\Sigma \cup $
\{-\} containing blank character '-' such that the string $a'$ is
derived from $a$ and string $b'$ is derived from $b$.
The alignment is denoted by\\
$$a{'}_{1} a^{'}_{2}\ldots a^{'}_{l}$$
$$b{'}_{1} b^{'}_{2} \ldots b^{'}_{l}$$\\
The length l of an alignment ($a^{'}$,$b^{'}$) is restricted to
$max \{m,n\}\le l \le m+n$, since column pairs  are not allowed.
In sequence alignment, the blank character '-' is referred as a
gap denoting insertion/deletion referred to as an \textit{indel}
(fig 2.7).
\begin{figure}
\begin{center}
\includegraphics[scale=0.45]{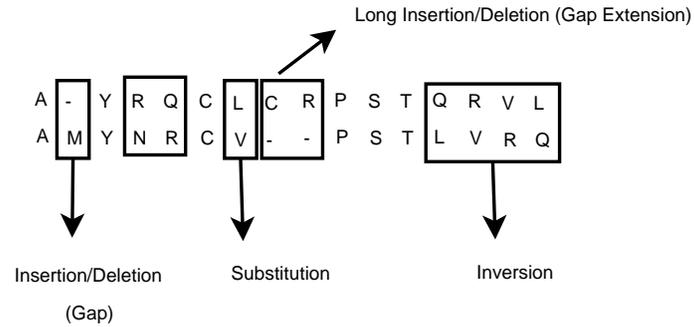}
\caption{Alignment between two sequences} \label{fig:2.7}
\end{center}
\end{figure}
\subsubsection{Alignment score}
\paragraph
\indent Scoring schemes are used to evaluate the alignment between
sequences. There are two scores that are used in alignment
evaluation:
 \begin{enumerate}
    \item substitution score
    \item insertion and deletion score.
\end{enumerate}
\subsubsection{Substitution score}
\paragraph
\indent Substitution scores are matrices developed based on
experimental data that encode the expected evolutionary change at
the amino acid level. One of the widely used substitution scores
in amino acid alignment is Point Accepted Mutation Matrix (PAM)
developed by Dayhoff in 1978 \cite{Krane2003}. The PAM matrix $M$
contains the probability of amino acid $i$ replaced by amino acid
$k$ in a certain evolutionary time period \cite{Krane2003}. For
example, 1PAM represents, 1 substitution per 100 residues
therefore, $n$PAM is $n$ accepted substitutions in 100 residues
(i.e, probability that amino acid $i$ will be replaced by amino
acid $k$ in sequences separated by $n$PAMs of evolutionary
distance). 1PAM is generally used for closely related sequences
and higher PAM matrices are used for distantly related sequences
(highly divergent). 1PAM was obtained by calculating the
substitution probabilities based on 71 groups of sequences with $>
80 \%$ sequence identity \cite{Eidhammer2004}.
The entries of 1PAM matrix ${M^1}$ is calculated as\\
$$M_{ij} = \log{\frac{\frac{m_j * F_{ij}}{\Sigma_i F_{ij}}} {f_i}}$$\\
where, the relative mutability $m_j$ is the number of times the
amino acid $i$ is substituted, $F_{ij}$ is the number of times
amino acid $i$ is substituted by amino acid $j$ and $f_i$ is the
frequency of amino acid $i$. Once $M$ is known, the matrix $M^n$
gives the probability of any amino acid mutating to any other
amino acid in $n$PAM units. The PAM matrix $M^n$ for $n
> 1$ can be obtained by matrix multiplication of $M^1$.
$$M^2 = M^1 * M^1$$
$$M^3 = M^2 * M^1$$
$$\vdots$$
$$M^n = M^{(n-1)} * M^1$$
\paragraph
\indent Another substitution matrix widely used is BLOSUM (Blocks
Substitution Matrix), developed by Henikoff and Henikoff in 1992
based on known alignments of more diverse sequences
\cite{Eidhammer2004}. The matrix is based on the ungapped
alignment (block) from the sequence alignment. Like the PAM
matrix, different BLOSUM scoring matrices are obtained for
different evolutionary distances. For example, BLOSUM80 matrix
represents sequences with approximately $80\%$ identity in
sequence alignment.
\paragraph
\indent The relationship between the two substitution matrices is
given as, BLOSUM with low percentage corresponds to PAM with large
evolutionary distances (i.e PAM250 $\rightarrow$ BLOSUM45, PAM120
$\rightarrow$ BLOSUM80). Lower numbered BLOSUM matrices are
appropriate for more distantly related sequences and  lower
numbered PAM matrices are appropriate for more closely related
sequences.
\subsubsection{Insertion/deletion score}
\paragraph
\indent Insertion and deletion scores are calculated based on the
gap opens (single insertion/deletion) and gap extensions (long
insertion/deletion)(fig 2.7). Since long insertions and deletions
are expected less than single insertion and deletion, they are
penalized less.
\begin{figure}
\begin{center}
\includegraphics[scale=0.45]{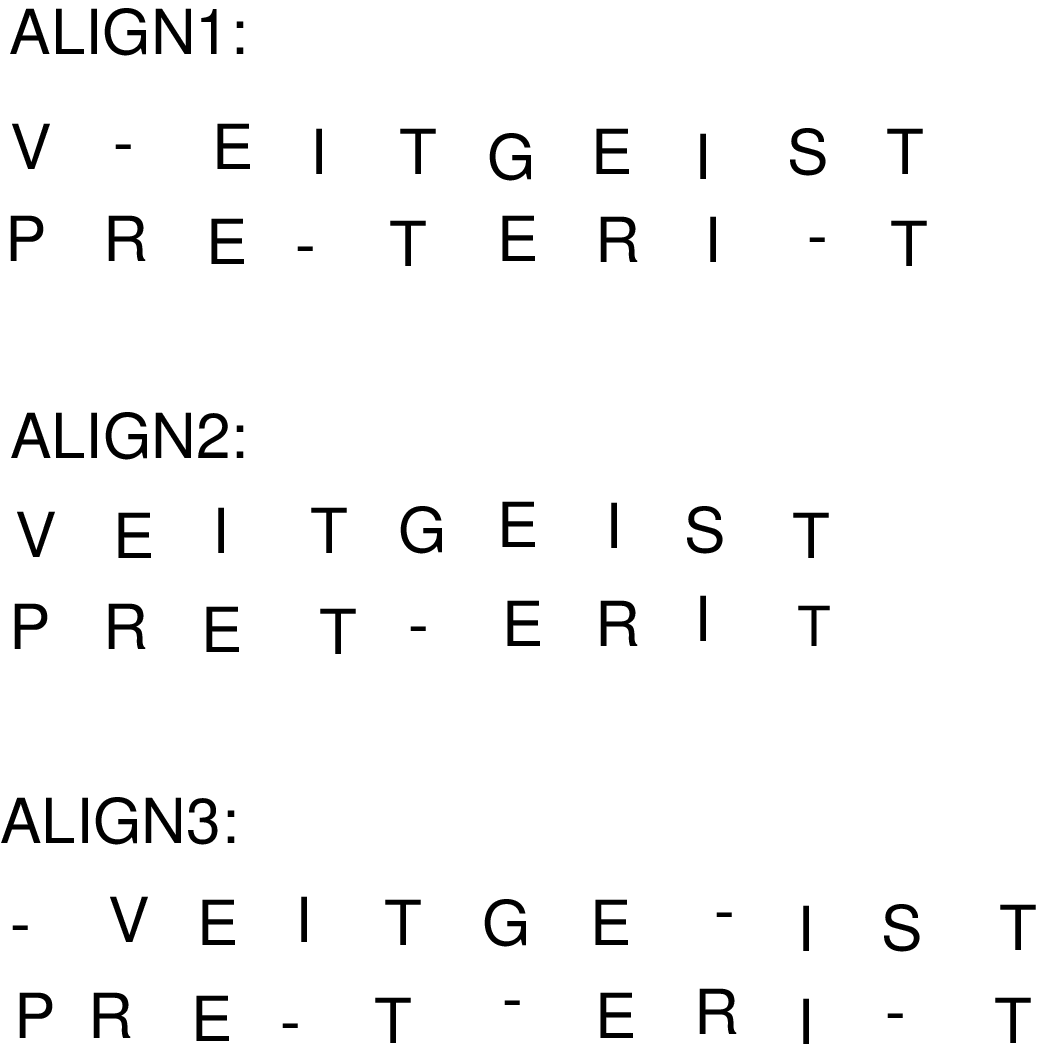}
\caption{Three possible alignments of two sequences}
\label{fig:2.8}
\end{center}
\end{figure}
\paragraph
\indent There could be more than one possible alignment between
the sequences (fig 2.8) but, the best alignment reflects the
evolutionary relationship between homologous sequences. In order
to find the best alignment, exhaustive search of all possible
alignments are not feasible. Therefore, alignment algorithms use a
dynamic program approach to break the problem into subproblems and
using partial results to compute the final answer
\cite{Krane2003}.
\subsubsection{Multiple sequence alignment}
\paragraph
\indent Multiple sequence alignment is an extension of pairwise
alignment to align more than two sequences simultaneously.
Multiple sequences are aligned in order to provide insight
into\begin{enumerate}
    \item characteristics of protein families
    \item identify motifs in sequences with a conserved biological
    function
    \item identify motifs of new proteins that would help to
    determine biological function.
\end{enumerate}
There are several multiple sequence alignment algorithms.
Algorithms with a heuristic approach are more commonly used than
the ones that give optimal alignment because optimal alignments
are practical only for a handful of sequences. Heuristic
algorithms are rapid, require less memory space and offer good
performance when used on relatively well conserved homologous
sequences. One of the most common heuristic approaches is
Progressive alignment used by ClustalW \cite{Zimmermann2003}. The
Progressive alignment algorithm works as follows:
\begin{enumerate}
    \item determine pairwise alignment between all pairs of
    sequences and their alignment scores.
    \item construct a guide tree (phylogenetic tree - see section ~\ref{phyl:phyltree}) using the
    alignment score.
    \item Align sequences according to the guide tree by aligning
    the most closely related sequences using sequence-sequence
    alignment first, then profile-sequence alignment( between
    an alignment and a sequence) and finally, profile-profile
    alignment( between alignments).
\end{enumerate}
In ClustalW substitution matrices and gap penalties vary at
different stages of alignment depending on the divergences of the
sequences to be aligned. Gap penalties depend on the substitution
matrices, the similarity of the sequences, and the length of the
sequences in order to introduce new gaps in the coil region rather
than in secondary structure regions \cite{Zimmermann2003}. One the
drawbacks of progressive alignment is that it is unreliable when
highly divergent sequences are aligned.
\subsubsection{Protein classification}
\paragraph
\indent A set of proteins that share a common evolutionary origin
reflected by their relatedness in function, which is usually
demonstrated by similarities in sequence, or in primary,
secondary, or tertiary structure is known as Protein
Family\cite{MCW}. Similarly, \textit{superfamiliy} is collection
of protein familes that have same overall domain structure (i.e,
same domain in same order) \cite{PIR}. There are several protein
family databases such as Prosite and Pfam. Proteins are also
classified based on the secondary structure similarities. Some
databases that group proteins based on structure classifications
are SCOP (Structural Classification of Proteins) and CATH (Class,
Architecture, Topology and Homologous superfamily).
\subsection{Phylogenetic tree}
\label{phyl:phyltree}
\paragraph
\indent Evolutionary relationships between species/sequences
(taxa) are called phylogenies and they are graphically represented
by trees known as \textit{Phylogenetic trees}. A phylogenetic tree
is made up of nodes and branches. Nodes represent distinct
taxonomical units. Nodes at the tips of the branches are
\textit{terminal nodes} and the internal nodes represent an
inferred common ancestor (fig 2.9). Branch lengths indicate the
amount of divergence between different species/sequences, longer
the lines between two species/sequences, the greater the
difference between them. Branch order refers to the genealogy of
the organism. If two species/sequences are closer to the branch
then closer their relationship.
\begin{figure}
\begin{center}
\includegraphics[scale=0.45]{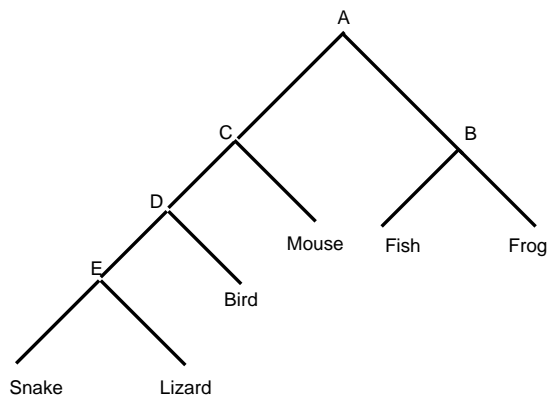}
\caption{Phylogenetic tree - (B,C,D,E) - internal nodes; snake,
lizard, bird, mouse, fish, frog - leaf nodes} \label{fig:2.9}
\end{center}
\end{figure}
 \subsubsection{Tree construction}
 \paragraph
 \indent One of the widely used tree constructions is the
 neighbor-joining Method based on distance matrices. The distance
 matrix consists of estimated distance between all pairs
 of taxas or operational taxon (OTU) (fig 2.10a) calculated from any method (say sequence alignment)
 used to find similarities and differences between sequences. The neighbor-joining method
 starts with a star tree having central node X of degree m ( number of neighbor of
 X). The new internal nodes are successively created and the degree of X
is reduced by 1 in each cycle. The iteration stops when the degree
of X becomes 3.
\begin{figure}[!hbp]
\begin{center}
\subfigure[Distance matrix of fig 2.11a]{\label{fig:2.10
a}\includegraphics[scale=0.40]{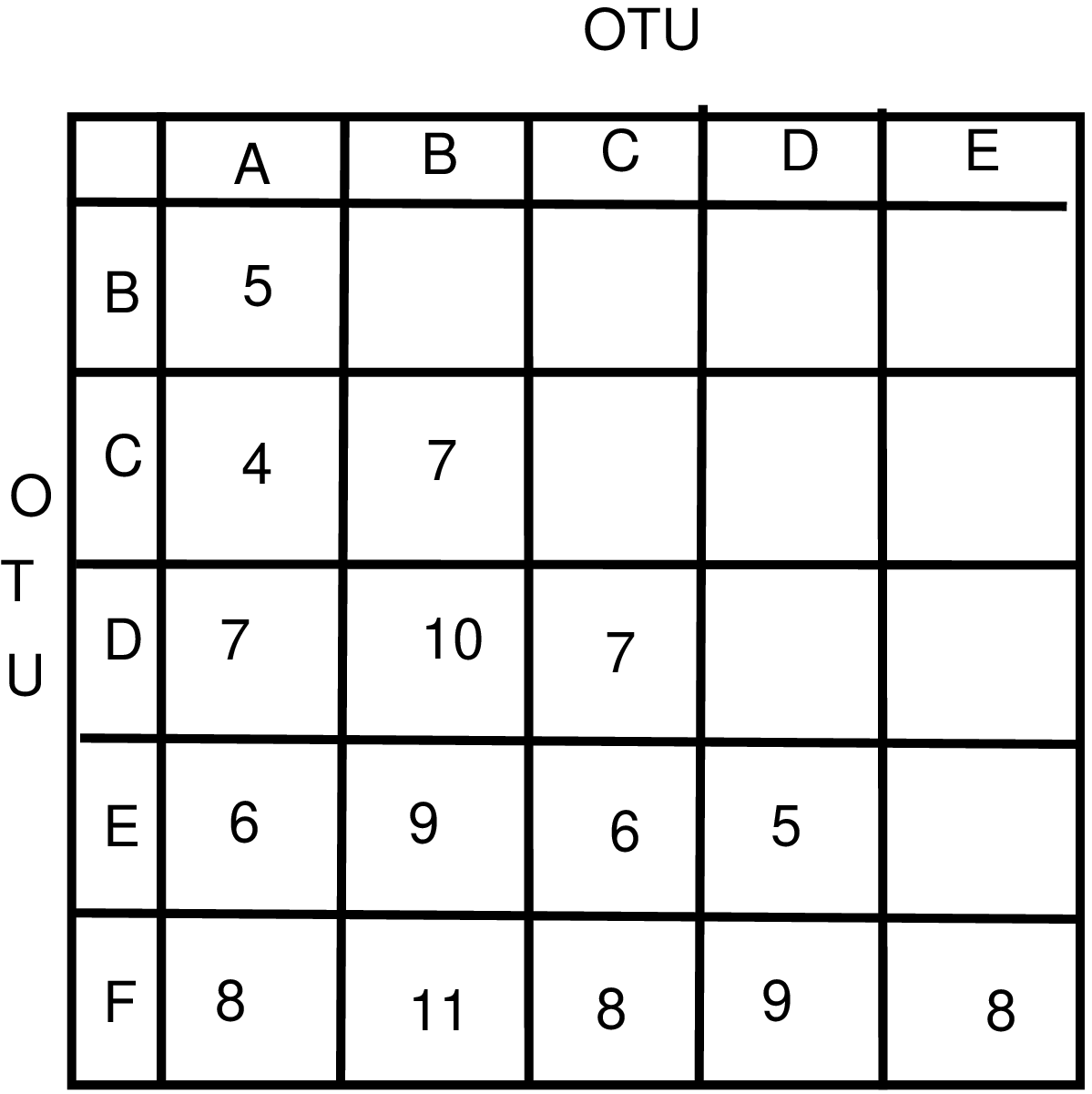}} \subfigure[(i) Distance
Matrix for fig 2.11a based on neighbor-joining method  (ii)
Distance matrix for fig 2.11b]{\label{fig:2.10
b}\includegraphics[scale=0.40]{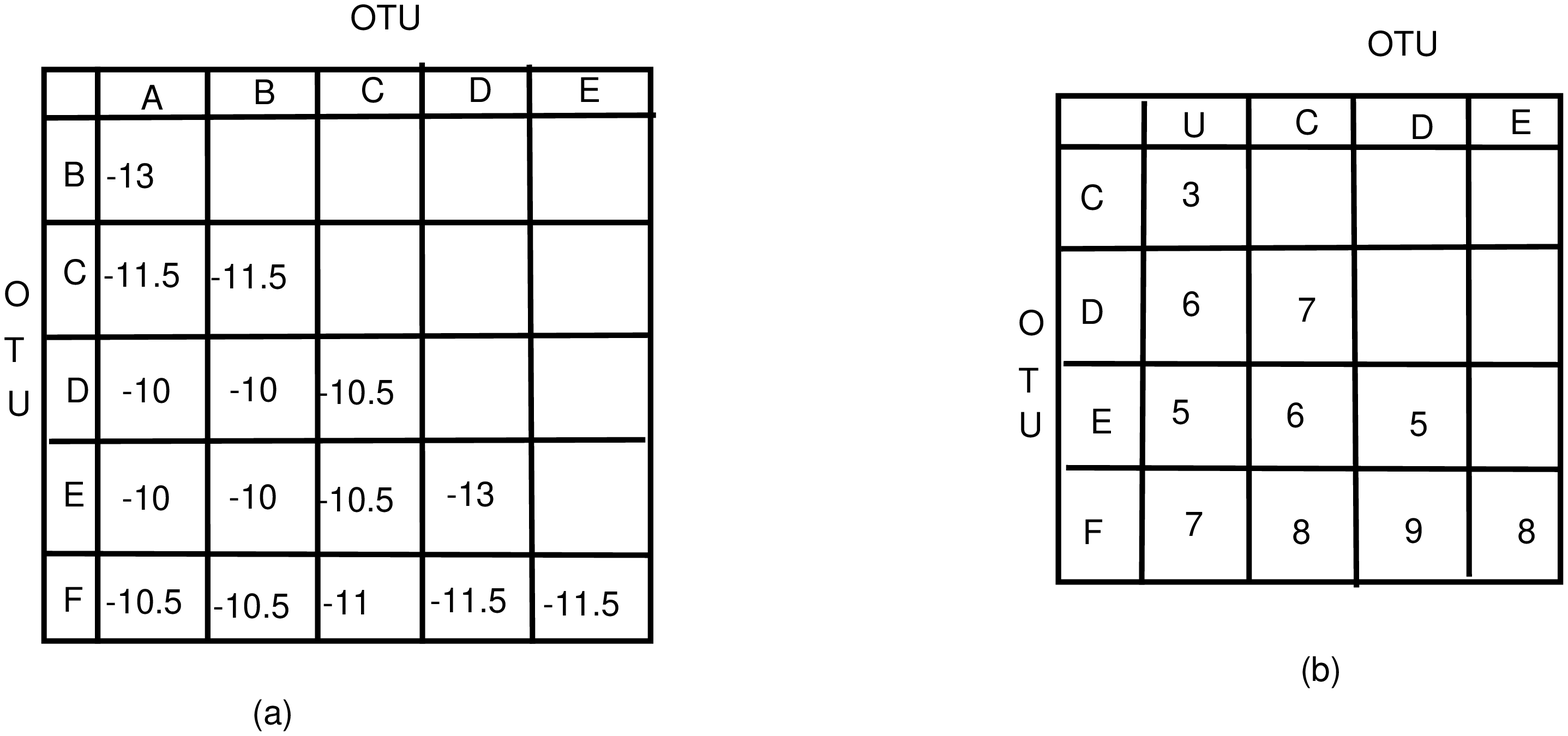}}
\end{center}
\caption{Distance matrices}\label{fig:2.10}
\end{figure}
\begin{figure}
\begin{center}
\includegraphics[scale=0.50]{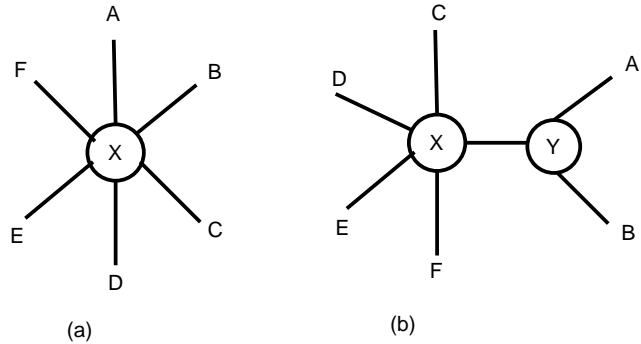}
\caption{(a) A star tree  (b) a tree with nodes A and B clustered}
\label{fig:2.11}
\end{center}
\end{figure}
\paragraph
 \indent The construction of phylogenetic tree  by neighbor-joining method (\cite{Saitou1987}) using distance matrix
 is explained using the following steps.
  \begin{enumerate}
    \item  start with a star tree (fig 2.11a).
    \item  calculate the net divergence $r(i)$ for each of the OTU's from all the
 OTU's using the distance matrix from fig 2.10a, net divergence for A is given as
 \begin{center}$r(A) = 5+4+7+6+8 = 30$ \end{center}
    \item  calculate a new distance matrix (fig 2.10b(i)) using the formula
 \begin{center} $M(ij) = d(ij)- [r(i)+r(j)] /(N-2)$ \end{center}
 where N is the number of OTU's, N=6 for fig 2.11a and $d(ij)$ is the distance between i and j, $d(AB) = 5$.
    \item  Choose two OTU's (A and B from fig:2.11a) from the distance matrix (fig:2.10b(i)) that
 has the smallest distance and create a new internal node Y that connects A, B and X.
    \item  calculate new branch length for A and B from Y using
    \begin{center} $S(AY) = d(AB)/2 - [r(A)-r(B)]/2(N-2), S(BY) =
    d(AB) - S(AY)$ \end{center} also, calculate the distance
    between Y to all the other nodes. The new distance matrix (fig 2.10 b(ii)) is
    created for fig 2.11b.
    \item  repeat process from step 2 until the degree of X
    becomes 3.
 \end{enumerate}

\chapter{Chaos and fractals for biological sequences}
\label{chap:chaosfractals}

\section{Fractal}
\paragraph
\indent A \textit{fractal} is a geometric figure that does not
become less complex  when you break it down into smaller and smaller
parts. This implies, a fractal is scale invariant. The word fractal
was coined by Mandelbrot from the Latin word \textit{fractus}
meaning broken or uneven, to describe objects that are too
irregular to fit into traditional geometry \cite{Falconer1990}.
For example, if we take a straight line and remove the middle
third from it, we obtain two small straight line segments and if
we continue this process repeatedly for smaller segments, in the
limit we obtain a fractal called the Cantor set (fig 3.1). Similar
examples are the Koch curve and the Sierpinski triangle (fig 3.2,
fig 3.3). In all of the above examples, the same structure has been
repeated at all scales. Therefore, these fractals are known as
\textit{self-similar} fractals. It is not necessary for fractals
to be self-similar. For example a coastline, the human body, or the sky on a
partly cloudy day are fractals without being self-similar\cite{Baranger}.
\begin{figure}
\begin{center}
\includegraphics{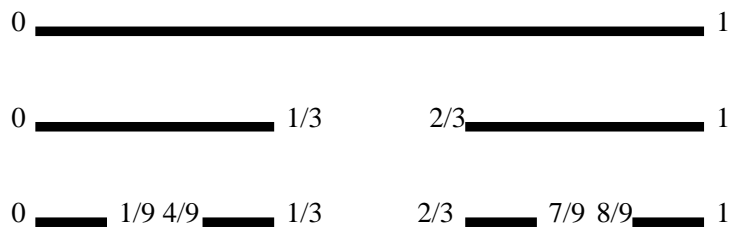}
\caption{The middle third Cantor set generated by repeated removal
of middle third of interval} \label{fig:3.1}
\end{center}
\end{figure}
\begin{figure}
\begin{center}
\includegraphics[scale=0.60]{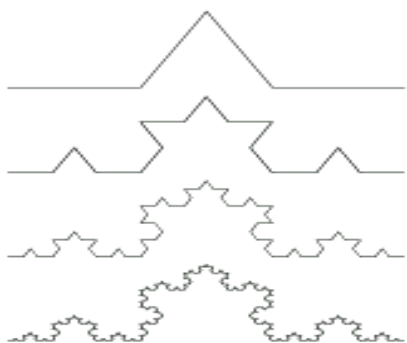}
\caption{The Koch curve generated by replacing the middle third of
each interval by the other two sides of an equilateral triangle
\cite{Fractal}:(Used)} \label{fig:3.2}
\end{center}
\end{figure}
\begin{figure}
\begin{center}
\includegraphics{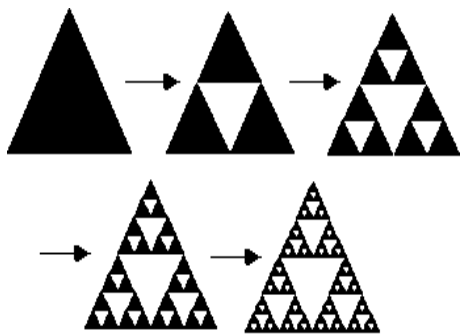}
\caption{The Sierpinski triangle generated by repeatedly removing
the inverted equilateral triangle from the center of the initial
equilateral triangle.} \label{fig:3.3}
\end{center}
\end{figure}
\subsection{Properties of fractals}\label{subsec:fractdim}
\paragraph
\indent Based on the examples above, we have the following
properties of fractals \cite{Falconer1990}. A fractal set F
\begin{enumerate}
    \item has a fine structure i.e detail on arbitrarily small
    scale;
    \item is too irregular to be described by traditional
    geometry;
    \item often has some form of self-similarity, perhaps
    approximate or statistical. For example, any part of the Cantor set F  in the interval
    $[0,\frac{1}{3}]$ and the interval $[\frac{2}{3},1]$ are geometrically similar to
    F. Figure 3.1, 3.2 and 3.3 contain copies of itself at
    different scales;
    \item is in most cases defined recursively. For example, the
    Cantor set is generated by repeatedly removing the middle
    third of intervals and the Sierpinski triangle is obtained by
    repeatedly removing the inverted triangle.
    \item usually has the fractal dimension (see \ref{subsec:fractdim}) greater than the
    topological dimension (see below) or Covering Dimension.
\end{enumerate}
 The topological dimension is defined as
    \\ \\
\indent  A covering of a subset $\mathcal{S}$ of a topological
space $\mathcal{X}$ is a collection $\mathcal{C}$ of open subsets
in $\mathcal{X}$ whose union contains all of $\mathcal{S}$\\ \\
\indent  A refinement of a covering $\mathcal{C}$ of S is another
$\mathcal{C}^\prime$ of $\mathcal{S}$ such that each set
$\mathcal{B}$ in $\mathcal{C}^\prime$ is contained in some set A
in $\mathcal{C}$. The idea is that the sets in
$\mathcal{C}^\prime$ are in some sense ``smaller'' than those in
$\mathcal{C}$ and provide a more finely detailed coverage of
$\mathcal{S}$.\\ \\\indent A topological space $\mathcal{X}$ has
topological dimension $m$ if every covering $\mathcal{C}$ of
$\mathcal{X}$ has a refinement $\mathcal{C}^\prime$ in which every
point of $\mathcal{X}$ occurs in at most $m+1$ sets in $\mathcal{C}^\prime$,
and $m$ is the smallest such integer.

\subsection{Mathematical fractals}
\paragraph
\indent Mathematically, a great variety of fractals could be
generated by iterating a collection of transformations, forming
what is known as an Iterated Function System (IFS). If all the
transformations in an IFS are \textit{contractive mappings} then
iterating these transformations would definitely converge to a
unique shape. A contractive mapping is a transformation $f$ that
reduces the distance between every pair of points. That is, there
is a number $s$ between 0 and 1 and
\\
$$dist(f(x,y),f(x',y')) \leq s * dist((x,y),(x',y'))$$
\\Formally, a contractive mapping, an IFS and an affine transformation are defined as
\paragraph
\indent \textbf{Definition 1:} A transformation $f$: $\mathcal{X}
\rightarrow \mathcal{X}$ on a metric space ($\mathcal{X},d$) is
called a contractive mapping if there is a constant 0 $\leq s \leq
1$ such that $d(f(x),f(y)) \leq s * d(x,y)$  $\forall  x,y
\varepsilon \mathcal{X}$, where $d$ is the Euclidean distance and
any $s$ the contraction factor of $f$.\\ \\
\indent \textbf{Definition 2:} Let $\mathcal{T}_1$,
$\mathcal{T}_2$, ..,$\mathcal{T}_N$ be a family of contractions on
$\Re^k$ and $\mathcal{S}$ be a closed bounded subset of $\Re^k$.
Then the system { $\mathcal{S}$: $\mathcal{T}$ = $U_{i=1}^n$
$\mathcal{T}_i$} is called an \textit{iterated function system}.\\
\\ \indent \textbf{Definition 3:} An \textit{affine transformation}
of $\Re^n$ is achieved by applying a linear transformation
followed by a translation. An affine transformation $\mathcal{T}$
of $\Re^n$ is represented in matrix-vector form as
\begin{center} $\mathcal{T}(x) = \mathcal{A}x +b,  x \in \Re^{n}$
and $\mathcal{A}$ is a transformation matrix \end{center}
\subsection{Example}
\paragraph
\indent The following example explains the mathematically
generated self-similar fractal called the Sierpinski triangle (fig
3.3). Consider $\mathcal{E}_{0}$ to be a unit triangle. The
contractive mapping for producing the
Sierpinski Triangle is given by three affine transformations. The three affine transformations for the Sierpinski triangle are\\
$$\mathcal{T}_{1}\left(\biggl[\begin{array}{c} x_{1} \\ x_{2} \end{array}\biggr]\right) = 1/2\biggl[\begin{array}{cc} 1 & 0 \\ 0 & 1 \end{array}\biggr] \hspace{0.4cm} \biggl[\begin {array}{c}x_{1}\\ x_{2}\end{array}\biggr] + \biggl[\begin{array}{c}0\\ 0\end{array}\biggr]$$
$$\mathcal{T}_{2}\left(\biggl[\begin{array}{c} x_{1} \\ x_{2} \end{array}\biggr]\right) = 1/2\biggl[\begin{array}{cc} 1 & 0 \\ 0 & 1 \end{array}\biggr] \hspace{0.2cm} \biggl[\begin {array}{c}x_{1}\\ x_{2}\end{array}\biggr] + \biggl[\begin{array}{c}1/2\\ 0\end{array}\biggr]$$
$$\mathcal{T}_{3}\left(\biggl[\begin{array}{c} x_{1} \\ x_{2} \end{array}\biggr]\right) = 1/2\biggl[\begin{array}{cc} 1 & 0 \\ 0 & 1 \end{array}\biggr] \hspace{0.2cm} \biggl[\begin {array}{c}x_{1}\\ x_{2}\end{array}\biggr] + \biggl[\begin{array}{c}1/4\\ \sqrt(3)/4\end{array}\biggr]$$
\\with a contractive factor $\frac{1}{2}$.
Applying $\mathcal{T}_{1}\left(\mathcal{E}_{0}\right)$,
$\mathcal{T}_{2}\left(\mathcal{E}_{0}\right)$ and
$\mathcal{T}_{3}\left(\mathcal{E}_{0}\right)$ to the triangle
$\mathcal{E}_{0}$ produces three smaller equilateral triangles
$\mathcal{E}_1$. Similarly, applying all the three transformations
to all the three vertices of each of the smaller triangles
$\mathcal{E}_1$ produces nine smaller triangles $\mathcal{E}_2$.
The iterative application of the three affine transformations
produces smaller and smaller triangles resulting in the Sierpinski
triangle (fig 3.3).\\  The iterative scheme is
\begin{center}$\mathcal{E}_0$ = a compact set\\
$\mathcal{E}_1 = \mathcal{T}(\mathcal{E}_0) =
\mathcal{T}_1(\mathcal{E}_0) \bigcup \mathcal{T}_2(\mathcal{E}_0)
\bigcup \mathcal{T}_3(\mathcal{E}_0)$ \\
$\mathcal{E}_2 = \mathcal{T}(\mathcal{E}_1) =
\mathcal{T}_1(\mathcal{E}_1) \bigcup \mathcal{T}_2(\mathcal{E}_1)
\bigcup
\mathcal{T}_3(\mathcal{E}_1)$\\
$\vdots$ \\
$\mathcal{E}_n = \mathcal{T}(\mathcal{E}_n-1) =
\mathcal{T}_1(\mathcal{E}_n-1) \bigcup
\mathcal{T}_2(\mathcal{E}_n-1) \bigcup
\mathcal{T}_3(\mathcal{E}_n-1)$\end{center} This sequence would
converge to a unique shape (the Sierpinski triangle) called an
\textit{attractor}. Since all the
transformations are applied in each step, this approach is a \textit{deterministic approach}. \\
\subsection{Fractal dimension}
\paragraph
\indent
 The \emph{dimension} is a topological measure of spacial extent. For example, a point has a dimension 0, a line has
 a dimension 1, a square has a dimension 2 and a cube has a dimension 3. However, topological dimension cannot
 be used to measure fractals, because, for example, when trying to measure the length of the Koch curve using line
 segments, as the number of line segments needed to measure the length increases, the length of the Koch curve
 increases, leading to infinity (fig 3.4). Table 3.1 lists the increasing length of the Koch curve as the
 number of line segments needed to measure the Koch curve increases (The initial line is of length 1). \\
\begin{table}
\begin{center}
 \begin{tabular}{|l|l|l|l|}
\hline n & length of the segment & number of segments &
Length of the Koch curve $\mathcal{L}_n$\\
\hline 0 & 1 & 1 & $\mathcal{L}_n = 1$\\
\hline 1 & $1/3$ & 4 & $\mathcal{L}_n = 4/3$\\
\hline 2 & $1/9 = 1/3^2$ & $16 = 4^2$ & $\mathcal{L}_n = 16/9 =
(4/3)^2$\\
\hline 3 & $1/27 = 1/3^3$ & $64 = 4^3$ & $\mathcal{L}_n
= 64/37 = (4/3)^3$\\ \hline \ldots & \ldots & \ldots & \ldots \\
\hline n & $1/3^n$ & $4^n$ & $\mathcal{L}_n = (4/3)^n$\\
\hline
\end{tabular}
\end{center}
\caption{Dimension of the Koch curve using length of line
segments} \label{Table: 3.1}
\end{table}
\\Similarly, when trying to compute the area of the Koch curve by covering it with triangles, as the
number of triangles needed to cover the Koch curve increases, the
area of the Koch curve decreases leading to zero (fig 3.4). The
initial triangle is an isosceles triangle with base 1 and height
 $\sqrt(3)/6$. In the next stage the Koch curve is covered with three smaller triangles whose
 base and height are reduced by $1/3$ compared to the initial triangle. As the process continues, at every
 stage the area of the triangles are reduced leading to zero. In the above metioned examples the dimension of the Koch curve leads to either infinity or zero producing no limiting value. Measuring an object in an dimension lower than the object produces infinity and higher than the object produced zero. This implies the dimension of the
 Koch curve is $>$ 1 but $<$ 2 a fractional value. Table 3.2 lists the area of the Koch curve for various
 sizes of the triangle needed to cover the Koch curve.
 \begin{table}
 \begin{center}
 \begin{tabular}{|l|l|l|l|}
\hline n & Area of the triangle & number of triangles &
Area of the Koch curve $\mathcal{A}_n$\\
\hline 0 & $\sqrt(3)/12$ & 1 & $\mathcal{A}_n = \sqrt(3)/12$\\
\hline 1 & $(\sqrt(3)/12) * (1/9)$ & 4 & $\mathcal{A}_n = (\sqrt(3)/12)*(4/9)$\\
\hline 2 & $(\sqrt(3)/12) * (1/81)$ & $16 = 4^2$ & $\mathcal{A}_n = (\sqrt(3)/12)*(16/81)$\\
\hline \ldots & \ldots & \ldots & \ldots \\
\hline n & $(\sqrt(3)/12) *(1/9)^n$ & $4^n$ & $\mathcal{A}_n = (\sqrt(3)/12)*(4/9)^n$\\
 \hline
\end{tabular}
\end{center}
\caption{Dimension of the Koch curve using area of triangles}
\label{Table: 3.2}
\end{table}
 \begin{figure}
\begin{center}
\includegraphics[scale=0.50]{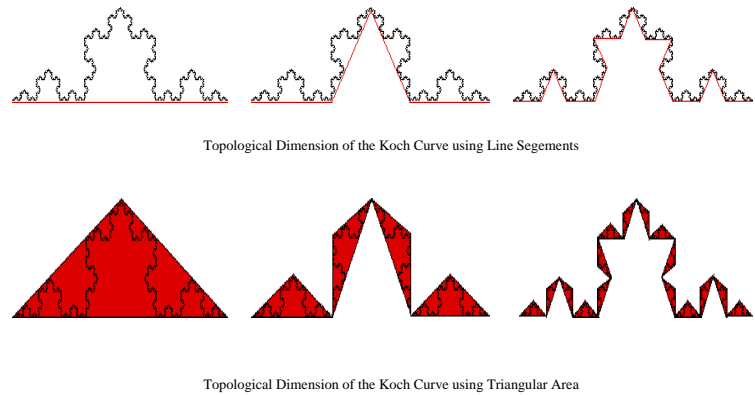}
\caption{Topological dimension of the Koch curve
\cite{Fractal}:(Adapted)} \label{fig:3.4}
\end{center}
\end{figure}
\paragraph
\indent
 Therefore, we use a better dimension called \textit{Box counting
 dimension} \cite{Fractal} to calculate the dimension of fractals. The box counting dimension of a fractal is
 calculated by covering the fractal with boxes and
 calculating the number of boxes $\mathcal{N}_{r}$ of size $r$ needed to cover the
 fractal (fig 3.5). The size of a fractal set is measured by its dimension($d_f$) given as:
\begin{center} $ d_f = \lim_{r \rightarrow 0}\frac{\log \mathcal{N}(r)}{\log r}$ \end{center}
For better approximation, the number of boxes needed to cover the fractal for various box sizes $r$ is calculated and the fractal dimension is the slope of log-log
 plot of size of the boxes against the number of such boxes needed to cover
 the fractal.
\begin{figure}
\begin{center}
\includegraphics[scale=0.55]{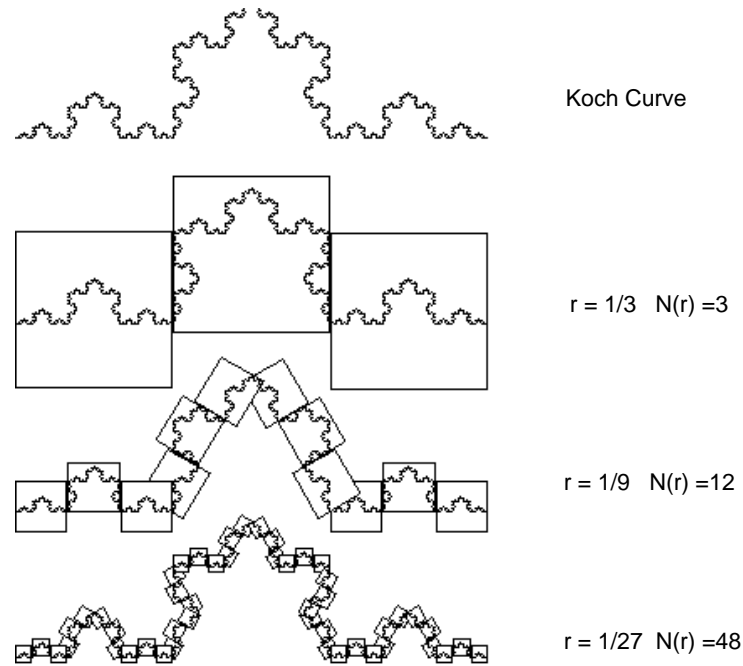}
\caption{Box counting dimension of the Koch curve
\cite{Fractal}:(Adapted), r - length of the sides and N(r) - no. of boxes needed to cover the fractal} \label{fig:3.5}
\end{center}
\end{figure}
\paragraph
\indent \textit{Example:} Figure 3.5 depicts the number of boxes
needed to cover the Koch curve for varying box sizes. The slope of
a log-log plot of size of the boxes against the number of boxes
needed to cover the Koch curve gives the fractal dimension of the
curve.
\section{Chaos game}
\paragraph
\indent Another approach to generate fractals is the random
approach of \textit{Chaos Game}. Consider a triangle and the three
transformations defined in section 3.1.3. Let the initial set be a
single point. Assume that, at each step, one of the three
transformations is randomly chosen and applied. Therefore, the
output at each stage is a single point. After some transient
behavior the points generated form a fractal - Sierpinski triangle. The iteration scheme is
\begin{center}$y_{0}$ = start point\end{center}
\begin{center}$y_{1} = \mathcal{T}_{1}(y_0)$ or $\mathcal{T}_{2}(y_0$) or $\mathcal{T}_{3}(y_0)$ \end{center}
\begin{center} $\vdots$ \end{center}
\begin{center}$y_{n} = \mathcal{T}_{1}(y_{n-1})$ or $\mathcal{T}_{2}(y_{n-1})$ or $\mathcal{T}_{3}(y_{n-1})$ \end{center}
\paragraph \indent For example, consider a triangle with vertices (0,0),(1,0) and
($1/2,\sqrt(3)/2$)(fig 3.6). The center of the triangle is chosen
as the starting point. Choose randomly one of the transformation,
say $\mathcal{T}_2$, and apply it to the center point: This would
produce a point (say $p$), which is the midpoint of the center and
the vertex (1,0). Again, randomly choose another transformation,
say $\mathcal{T}_3$, and apply it to the previously produced point
($p$): The new point produced is the midpoint of the point $p$ and
the vertex ($1/2,\sqrt(3)/2$). As this process is continued for
large number of times,  the image produced looks like a Sierpinski
triangle. The chaos game can be played with any number of vertices, four, five, six and so on. If the vertices are not selected uniformly at random in chaos game then various patterns are produced. This reveals some kind of order in the sequence.
\begin{figure}
\begin{center}
\includegraphics[scale=0.65]{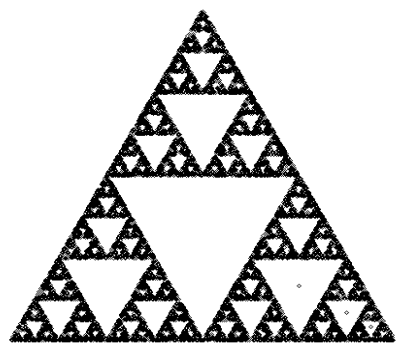}
\caption{Sierpinski triangle using the chaos Game} \label{fig:3.6}
\end{center}
\end{figure}
\paragraph
\indent Intuitively, order (non-randomness) means the sequence has a
structure. Therefore, the chaos game can be a used as a tool to study the non-randomness of any sequence visually.
If the chaos game can be extended to play on DNA or protein sequences then various patterns/structure in them could be revealed.

\chapter{Chaos game representation of DNA and protein sequences}
\label{chap:DNAPro}

\section{ Chaos game representation of DNA sequences}
\paragraph
\indent Chaos game representation (or CGR) is a visual
representation technique used, among others to study the patterns
in gene structures \cite{Jeffrey1990}. The chaos game is played on
a square using IFS. The nucleotide bases (A,C,G,T) correspond to
the four vertices; the first point is plotted halfway between the
center of the square and the corresponding vertex of the first
nucleotide base in the sequence, and each subsequent point is
plotted halfway between the previous point and the vertex of the
subsequent nucleotide base from the sequence (fig 4.1a). The chaos
game when applied to DNA sequences showed fractal structures (fig
4.1b , fig 4.1c). Figure 4.1b represents the attractor of chaos
game of (\textit{Human Beta Globin-HUMBB}). The 'double
scoop'(sparse regions on fig 4.1b) structure in the attractor is
due to the paucity of points in various regions of the square
\cite{Jeffrey1990}. Similarly, when A $\&$ T and C $\&$ G were
plotted opposite to one another, the paucity of points is
represented by squares instead of 'double scoop' (fig 4.1c). The
patterns are repeated on various scales in the attractor
exhibiting the property of self-similarity. These patterns
revealed the non-random nucleotide composition in DNA sequences.
In a CGR, any $ith$ point in the attractor uniquely represents the
$ith$ long initial subsequence of the sequence. The attractor
depicts the base composition of the gene sequence
\cite{Jeffrey1990}; the square, when divided into four, sixteen,
sixty four sub-quadrants and so on represents the mono, di- and
tri- etc. nucleotides subsequences (fig 4.1d).
\begin{figure}[!hbp]
\subfigure[CGR-AGAAT; Plotting "AGAAT"- 'A' is plotted half-way
between the center and the vertex representing A, 'G' is plotted
half-way between the previous point plotted and the vertex
representing G, 'A' is plotted half-way between the previous point
and the vertex representing A, next 'A' is plotted halfway between
the previous point  and the vertex representing A and similarly
'T' is
plotted]{\label{fig:4.1-a}\includegraphics[scale=0.55]{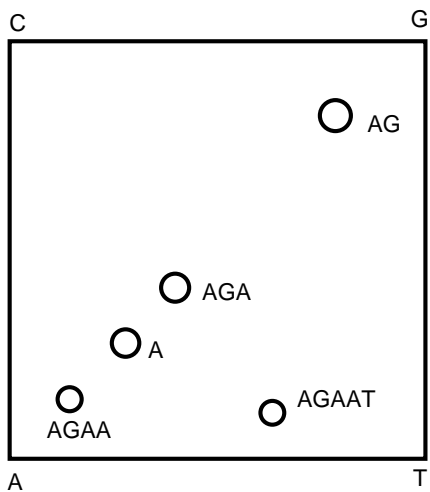}}
\subfigure[CGR-HumanBetaGlobin; Nucleotides A,C,G,T are
represented as red, violet, green and blue; paucity of
dinucleotide CG forms a 'Double Scoop'
pattern]{\label{fig:4.1-b}\includegraphics[scale=0.65]{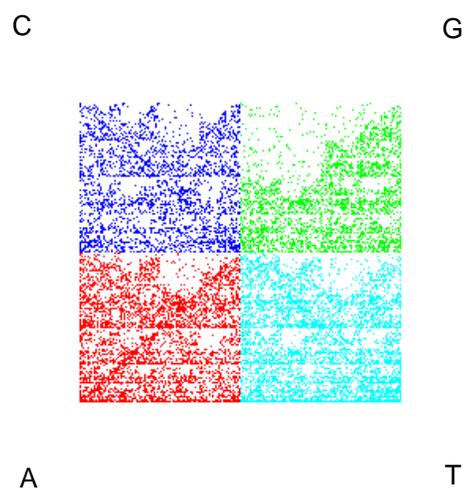}}
\subfigure[CGR-HumanBetaGlobin; paucity of dinucleotide CG forms a
square pattern as vertices C and G are opposite to each
other.]{\label{fig:4.1-c}\includegraphics[scale=0.65]{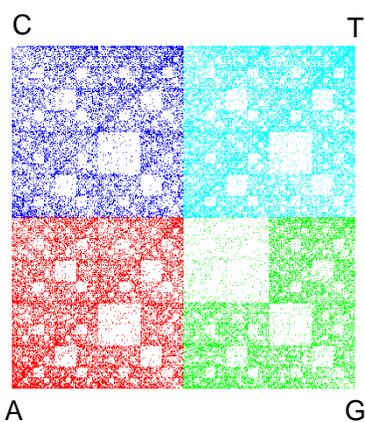}}
\subfigure[Mono-, di-, and Tri- nucleotide
configuration]{\label{fig:4.1-d}\includegraphics[scale=0.65]{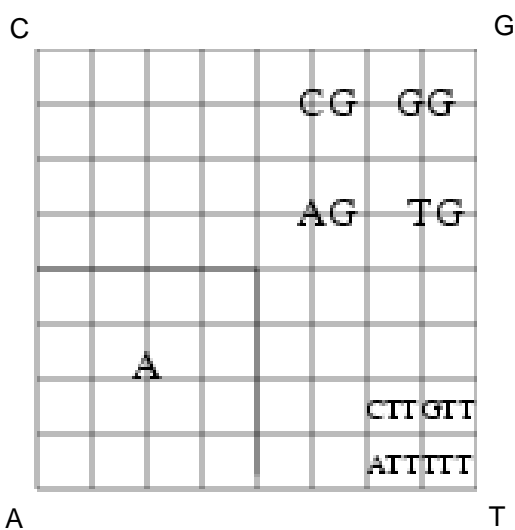}}
\caption {Properties of CGR} \label{fig:4.1}
\end{figure}
\paragraph
\indent Using a CGR, the presence or absence of a sequence of
nucleotides in any DNA sequence can be mathematically
characterized \cite{Dutta1992}. Dutta et.al \cite{Dutta1992} gave algorithms to find the
subsequences corresponding to any given point in CGR, and to
simulate CGR patterns of a sequence by predicting the order of
nucleotides using the probability of occurrences of di or tri
nucleotides \cite{Dutta1992}. These algorithms are presented in the
appendices A and B.
\paragraph
\indent Hill et.al in 1992 \cite{Hill1992} examined the coding region of the CGR's of seven
\textit{globin} genes from human and the CGRs of 29 closely related \textit{alcohol
dehydrogenase} genes from phylogenetically divergent species. CGRs of Human \textit{globin} coding
regions and the CGR of the entire Human Globin gene (coding and
non-coding) are visually similar to each other but the
self-similarity was not readily visible in the individual sequences
due to smaller number of points (2000 nucleotides). Also, the
di-nucleotide frequencies of the \textit{globin} genes from human
are not significantly different from one another. Therefore, the
di-nucleotide frequencies partially accounted for the
self-similarity in the CGR patterns \cite{Hill1992}. CGRs from
coding regions of \textit{alcohol dehydrogenase} gene from the
same species (ADH1, ADH2 etc) are similar to one another. Also,
ADH CGRs of closely related species such as human, rodent, primate
were similar to one another\cite{Hill1992}. CGRs of unrelated
genes from the same species are more similar to one another than
sequences from unrelated species. Therefore,  Hill et.al in 1992 \cite{Hill1992} said that
the CGR patterns reflect genome type specificity which could be
the result of mutation rates of  mono, di, tri nucleotide bases
and so on and said the evolution of a gene and its coding sequences
should not be examined in isolation, genome specific differential
mutation in di-nucleotides or oligonucleotide should be taken into
account \cite{Hill1992}. Hill et.al in 1997 \cite{Hill1997} studied 28 complete mitochondrial genomes using CGR. They said, the global DNA sequence organizationof mitochondrial genomes is species-type specific. The species-type specific patterns appear primarily due to the dinucleotide composition.
\paragraph
\indent CGRs generated from simulated sequences using a
first-order Markov-chain probability matrix for \textit{Human Beta
Globin} and second-order Markov-chain probability matrix for
\textit{Bacteriophage Lambda} were similar to CGRs of the original
Human Beta Globin and Bacteriophage Lambda sequences \cite{Goldman1993}. The first-order and second-order
Markov probability matrices were obtained from calculating the
dinucleotide and trinucleotide frequencies directly from the DNA
sequences without reference to the CGR. Therefore Goldman in 1993
\cite{Goldman1993}, suggested that CGR is a particular case of
Markov-Chain model and CGR is only limited to represent mono, di
and tri nucleotide representation of the sequences
\cite{Goldman1993}.
\paragraph
\indent The Markov chain model is limited to produce only integer
number of bases whereas the frequency matrix obtained from CGR
(FCGR) can produce non-integer number of bases \cite{Almedia2001}
in contradiction to statement by Goldman in 1993 \cite{Goldman1993}. The frequency matrix for
oligonucleotides of length $n_{c}$ is obtained by dividing the CGR
into a $2^{n_{c}} * 2^{n_{c}}$ grid. Then the Markov chain
probability matrix could be obtained from
FCGR only if the quadrant $k$ satisfies the condition in the following equation to produce an integer order\\
\begin{center} $k = 2^{2{n_{c}}} , n_{c}$ is an integer $ \ge 1$ \end{center}
But, if the condition '$n_{c}$ is an integer $\ge 1$' is removed
then
$$n_{c} = \log_{2}(k) /2,$$\\
i.e FCGR can track the frequency of oligonucleotide of non-integer
order. Therefore, CGR enables the determination of the frequency
of redundant fractionary sequences also, FCGR of non-integer order
can be used to calculate global distance and local similarities
between sequences \cite{Almedia2001}.
\paragraph
\indent If $k$=3, then the patterns in CGR are determined by
mononucleotide, dinucleotide and trinucleotide frequencies but if
$k > 3$, then longer oligonucleotides may influence the CGR
patterns. Therefore, Wang et.al \cite{Wang2004} said, a CGR of 1/$2^k$ resolution is completely
determined by all the frequencies of length $k$ when the length of
the DNA sequence is longer than $k$ . They also analyzed the relationship between dinucleotide
relative abundance profile \footnote{ratio of the dinucleotide
frequency to the frequency of two single nucleotide composing this
dinucleotide}(DRAP) and CGR. DRAP can be computed from second-order or
dinucleotide frequency FCGR, but second order FCGR cannot be
computed from DRAP. Therefore, DRAP or rFCGR (relative FCGR) is a
special case of FCGR.  Wang et.al \cite{Wang2004} said, an $n$
th order FCGR provides more info than DRAP. However, the
second-order FCGR or DRAP is a good choice of genomic signature \footnote{The whole set of short oligonucleotide frequencies observed in a DNA sequence is species-specific and is thus considered as a GENOMIC SIGNATURE} as
the computational cost is higher for higher order FCGRs
\cite{Wang2004}. A new distance measure called \textit{image
distance} used to calculate the distance between genomic
signatures of two DNA sequences \cite{Wang2004} is given by\\
\begin{center} $dI_\mathcal{R}\left(\bar{\mathcal{A}},\bar{\mathcal{B}}\right) = 1/4^k * \sum_{i=1}^{2^k}\sum_{j=1}^{2^k} \left| density_\mathcal{R}\left(\bar{\mathcal{A}}\right)_{i,j}- density_{\mathcal{R}}\left(\bar{\mathcal{B}}\right)_{i,j}\right|$\end{center}
\begin{center} $\bar{\mathcal{A}} = 4^k/\sum_i\sum_j a_{i,j} * \mathcal{A}$ and \end{center}
\begin{center} $\bar{\mathcal{B}} = 4^k/\sum_i\sum_j a_{i,j} * \mathcal{B}$ \end{center}
where $\mathcal{A}$ and $\mathcal{B}$ are frequency matrices of
$k$th order, $\mathcal{R}$ is the radius of the neighborhood
centered at $(i,j)$ and \textit{$density_\mathcal{R}$}. The phylogenetic trees built using the Euclid
distance, Pearson distance and Image distance between two CGRs have
proven to be more compatible with phylogenetic relatedness of
species than the tree obtained from ClustalW \cite{Wang2004}.
\section{Chaos game representation of protein sequences}
\label{sec:CGRPro}
\subsection{Chaos game using a 20 sided polygon}
\paragraph
\indent The chaos game representation of protein sequences was
used to find the motifs in the sequence, describe regularities in
structure elements, and evaluate various secondary structure
prediction algorithms \cite{Fiser1994}.
\subsubsection{Method}
\indent Fiser et.al in 1994 \cite{Fiser1994} applied Chaos Game
Representation to protein sequences to investigate the motifs in
the protein database and protein sequences. A 20-sided regular
polygon was used to represent the 20 different amino acids. The
$(x,y)$ coordinates of
each of the vertices were given as\\
$$v_{i,x}   = cos(2\pi* i/n)$$
$$v_{i,y} = sin(2\pi *i/n)$$
\subsubsection{Plotting}
\paragraph
\indent The coordinates of the 0th point are [0,0] and the $m$th
point was given by dividing the distance
 between the $(m-1)$th point and the vertex representing the $m$th amino acid using the dividing ratio
 $s_1$ and $s_2$. \\ The coordinates of the points are\\
 $$p_{m,x}  = (v_{m,x} - p_{m-1,x}) * s_{2}+ p_{m-1,x }$$
 $$p_{m,y} = (v_{m,x} - p_{m-1,y}) * s_{2} + p_{m-1,y}$$\\
The dividing ratio $s_1$: $s_2$ is 0.135:0.865 calculated
from\\
$$s_{1}  = sin(2\pi * i/n) / ( 1 + sin(2\pi * i/n))$$
$$s_{2} = 1 / (1 + sin(2\pi * i/n))$$ \\
A lower dividing ratio was used in order to obtain an unambiguous
and decodable fractal for an attractor.
\begin{figure}[!hbp]
\begin{center}
\includegraphics[scale=0.75]{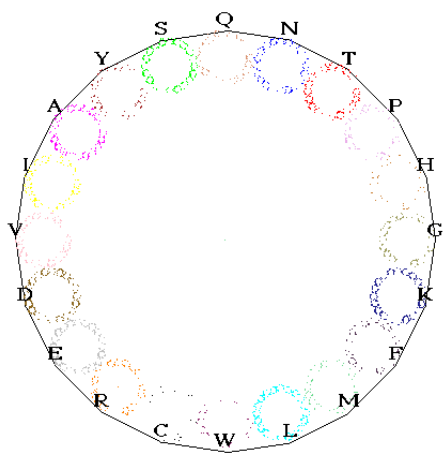}
\caption{CGR using 20-gon of DNA Polymerase Human Alpha Chain:
Length = 1462} \label{fig:4.4 Motif Detection}
\end{center}
\end{figure}
\subsubsection{Properties}
\paragraph
\indent The attractor produced was a 20-gon in which there are
small separate 20-gons at every vertex. The small 20-gon can
further contain smaller 20-gons in their vertices etc (fig 4.2).
For example, an amino acid subsequence IDEAL can be decoded by zooming the
20-gon at vertex L followed by the  20-gon at vertex A of the L
polygon, the 20-gon at vertex E of the A polygon, the 20-gon at
vertex D polygon and the 20-gon at vertex I of the D polygon.
Theoretically, each point represents the preceding sequence motif.
Eventhough the attractor can be used for identifying subsequences
or motifs, they are indistinguishable as the sequence length
increases. Fiser et.al extended
the chaos game to study the regularities in secondary structure
elements of proteins. The major secondary structures helix, sheet
and turn were represented as vertices of a triangle and the random
coil as the center. The attractor produced was used to study the
frequency of attachment of various secondary structure and
evaluate structure prediction methods. Therefore, CGR could be
used to study both primary and 3D structures of proteins
\cite{Fiser1994}.
\subsubsection{Limitation}
\paragraph
\indent The major drawback of this approach is that as the
sequence length becomes larger all the polygons looks equally
filled. Therefore, various sequence motifs become
indistinguishable.
\subsection{Chaos game using a rectangle}
\subsubsection{Method}
\paragraph
\indent A rectangle was used to represent the sequentiality and composition
of amino acids in a sequence. The rectangle was divided into 5 x 4
sub-rectangles representing 20 different amino acids.
\begin{figure}[!hbp]
\begin{center}
\includegraphics[scale=0.75]{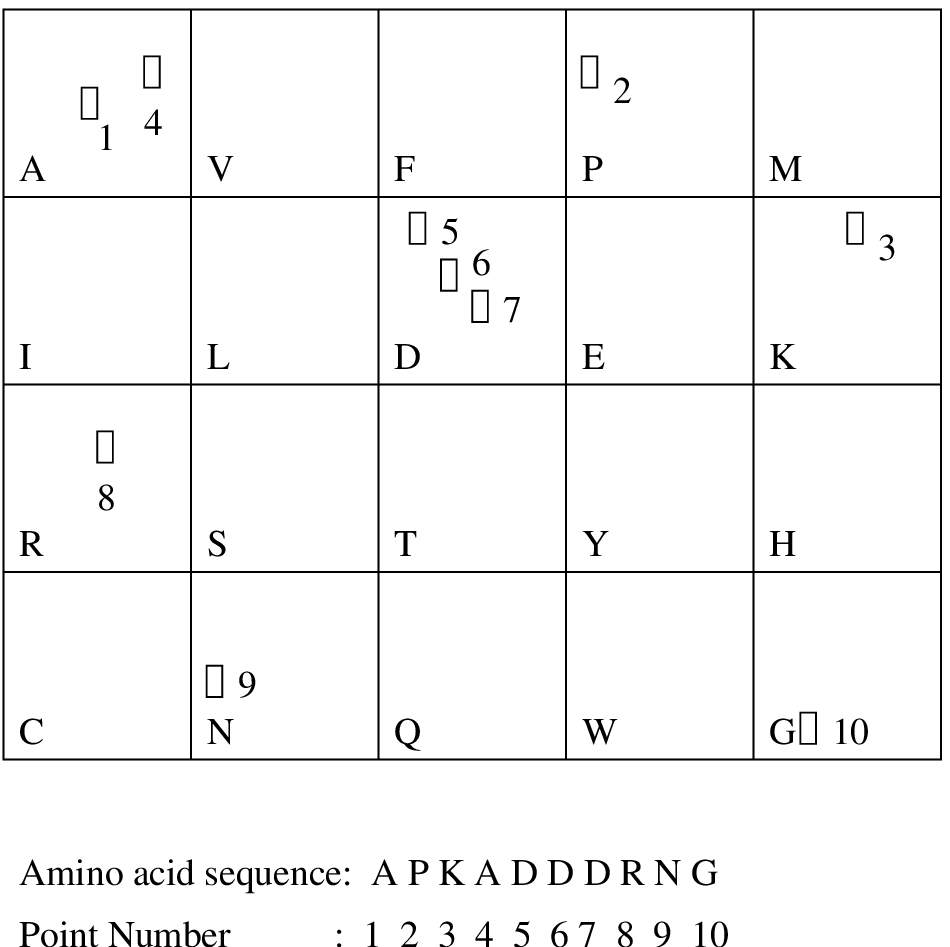}
\caption{2D point representation using rectangle for sequentiality
and composition of amino acids} \label{fig:4.5 Point Pattern}
\end{center}
\end{figure}
\subsubsection{Plotting}
\paragraph
\indent The chaos game is played as follows: the first point is
plotted in the middle of the sub-rectangle labelled with the first
amino acid in the sequence. The \textit{ith} point is plotted by
scaling the \textit{(i-1)th} point by 1/5 in $x$ -direction and
1/4 in $y$-direction and moving the point to the sub-rectangle
labelled with the \textit{ith} amino acid (fig 4.3).
\subsubsection{Properties}
\paragraph
\indent Some characteristic properties noted were that the points
that follow after the insertion and deletion of amino acids are
shifted but the degree of shifting is reduced by $5^n$ in the
$x$-direction and $4^n$ in the $y$-direction, where $n$ is the
number of letters after the inserted/deleted position. Therefore, insertion/deletion does not change the
overall visual impression of the point pattern
. The reduction in shift was also noted when there were repeats in amino acids.
\cite{Pleibner1997}.
\subsection{Chaos game using a 12 sided polygon}
\subsubsection{Method}
\paragraph
\indent CGR can be used to study characteristic patterns of
protein families \cite{Basu1997}. A 12 sided regular polygon was
used to plot a concatenated amino acid sequence of proteins from
protein family (fig 4.4). Each vertex of the polygon corresponded
to a group of amino acid residues of conservative substitutions
(section 3.3) and the amino acids along the vertices of the
polygon were placed in the order of decreasing normalized
hydrophobicity.
\subsubsection{Plotting}
\paragraph
\indent The chaos game was played as in the case of DNA sequences
using the square \cite{Jeffrey1990}. A grid counting algorithm was
used to quantify the CGR.  The 12 sided regular polygon was
divided into 24 segments and the number of points in each segment
was calculated, the percentage of points in each segment was given
by $C_{j}$ / N* 100(j=1..24).
\subsubsection{Properties}
\paragraph
\indent Even though no fractals were detected in the CGRs, there
were specific statistical biases in the distribution of different
amino acids, mono-, di-, tri- or higher order peptides in
functional classes of proteins \cite{Basu1997}. The CGRs of a
protein family were dependent on the relative order of the
residues. The patterns were insensitive to the shuffling of less
abundant residues along the vertices of the polygon, but
sensitive to the shuffling of more abundant residues along the
vertices. The plots are visually similar for a protein class for
many different orientations of the residues. The grid count was
also invariant for a particular family of proteins and particular
orientation of the residue group along the vertices of the CGR
irrespective of the number of sequences concatenated and the order
of concatenation. Therefore, the grid count can be used as a
diagnostic signature of a protein family for identifying new
members of the family and CGR has a potential to reveal
evolutionary and functional relationship between proteins having
no significant homology \cite{Basu1997}.
\begin{figure}
\begin{center}
\includegraphics[scale=0.60]{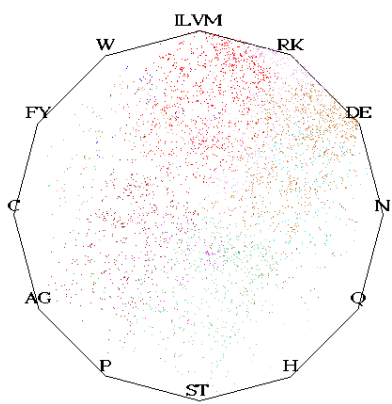}
\caption{CGR using 12 sided polygon; HEAT SHOCK PROTEIN 90 (hsp90)
family} \label{fig:4.6}
\end{center}
\end{figure}
\subsection{Chaos game using a square}
\paragraph
\indent Multi-fractal and correlation analysis were performed on
CGR of bacteria families to study the phylogenetic relationship
between sequences and sub-families \cite{Yu2004}.
\subsubsection{Method}
\indent A detailed hydrophobic or non-polar and hydrophilic or
polar (HP) model is used to represent the four classes of amino
acids - non-polar, uncharged, positively charged and negatively
charged as four vertices of the square.
\subsubsection{Plotting}
\paragraph
\indent The chaos game is played as in the case of DNA sequences
\cite{Jeffrey1990}. The square is divided into equal meshes and a
measure $\mu$ for each mesh was calculated by dividing the number
of points lying in the subset of the CGR  by the length of the
sequence. This is represented as a measure matrix $\mathcal{A}$.
Also, a symbolic sequence is created based on the probability of
amino acids in the position of the original sequence and a measure
matrix is calculated for it and referred to as a measure of
fractal background $\mathcal{A}^{f}$. A new measure matrix
$\mathcal{A}^{d}$ is obtained by subtracting the measure matrix of
the fractal background from the measure matrix of the original
sequence and used for calculating the correlation distance.\\
\subsubsection{Properties}
\paragraph
\indent The phylogenetic tree based on correlation distance is
more precise as the mesh size increases \cite{Yu2004}.
Multifractal analysis (\cite{Harte2001}) of the test sequences
exhibited multi-fractal like forms indicating that the protein
sequences of a complete genome are non-random. Also, said, the
correlation analysis is more precise than the multi-fractal analysis for the phylogenetic problem\cite{Yu2004}.
\subsection{Summary}
\paragraph
\indent The following table summarizes the results of Chaos Game
Representation on Protein Sequences in two dimensions.
\begin{landscape}
\begin{table}[!hbp]
\begin{tabular}{|p{3in}|p{2.5in}|p{1.5in}|p{0.5in}|p{0.3in}|}
\hline
 Approach & Novel Advances& Limitations & Authors & Year\\ \hline
Chaos Game using 20 sided polygon, 20 vertices represents 20
amino acids & motif detection in protein database,regularities in
secondary structure elements and evaluation of secondary structure
prediction methods& for large sequences, motif detection is not easy due to resolution of the monitor& Fiser et.al & 1994\\
\hline
Chaos Game using a 5 x 4 rectangle &sequentiality and composition of amino acids & & Pleibner & 1997\\
\hline Chaos Game using 12 sided polygon, 12 vertices represent
12 groups of amino acids & Characteristic patterns of Protein family, measure to detect protein family for a given protein & individual representation of amino acid is lost & Basu et.al & 1997\\
\hline Chaos Game using a square, vertices represent non-polar,
uncharged, positively charged and negatively charged amino acids
&evaluate phylogenetic tree of bacteria &individual representation of amino acid is lost & Yu et.al & 2004\\
\hline
\end{tabular}
\caption{Summary - Chaos game representation of protein sequences
in two dimension} \label{Table: 4.1}
\end{table}
\end{landscape}

\chapter{Three dimensional CGR of protein sequences}
\label{chap:3DCGR}

\paragraph
\indent The chaos game representation of proteins in two
dimensions discussed in the previous chapter helped to identify
motifs in the protein databases and to test secondary structure
prediction methods \cite{Fiser1994}, reveal patterns that
distinguish protein families \cite{Basu1997}, understand the
sequentiality and composition of amino acids \cite{Pleibner1997}
and better understanding of the bacterial family homology
\cite{Yu2004}. In this chapter, a new three dimensional approach
to CGR (3D-CGR) as an analysis tool of protein sequence is
proposed, with the following objectives:
\begin{itemize}
 \item   use the three dimensional approach to detect protein homology
 \item   assess the impact of dinucleotide bias at the amino acid level on 3D-CGR derived protein homology and
 \item   use the three dimensional approach to detect shuffled motifs.
\end{itemize}
\section{Three dimensional structure and amino acid mapping}
\paragraph
\indent In order to play the chaos game for protein sequences in
three dimensions an \textit{icosahedron} was chosen to be the
geometric model. An icosahedron is a geometric solid that has
twelve vertices, thirty edges and twenty faces. An icosahedron was
chosen because the  twenty amino acids of a protein can be
represented by the twenty faces of an icosahedron.
\begin{figure}[b]
\begin{center}
\subfigure[2D Representation; Amino acid K represents the face at
the
back]{\label{fig:5.1-a}\includegraphics[scale=0.40]{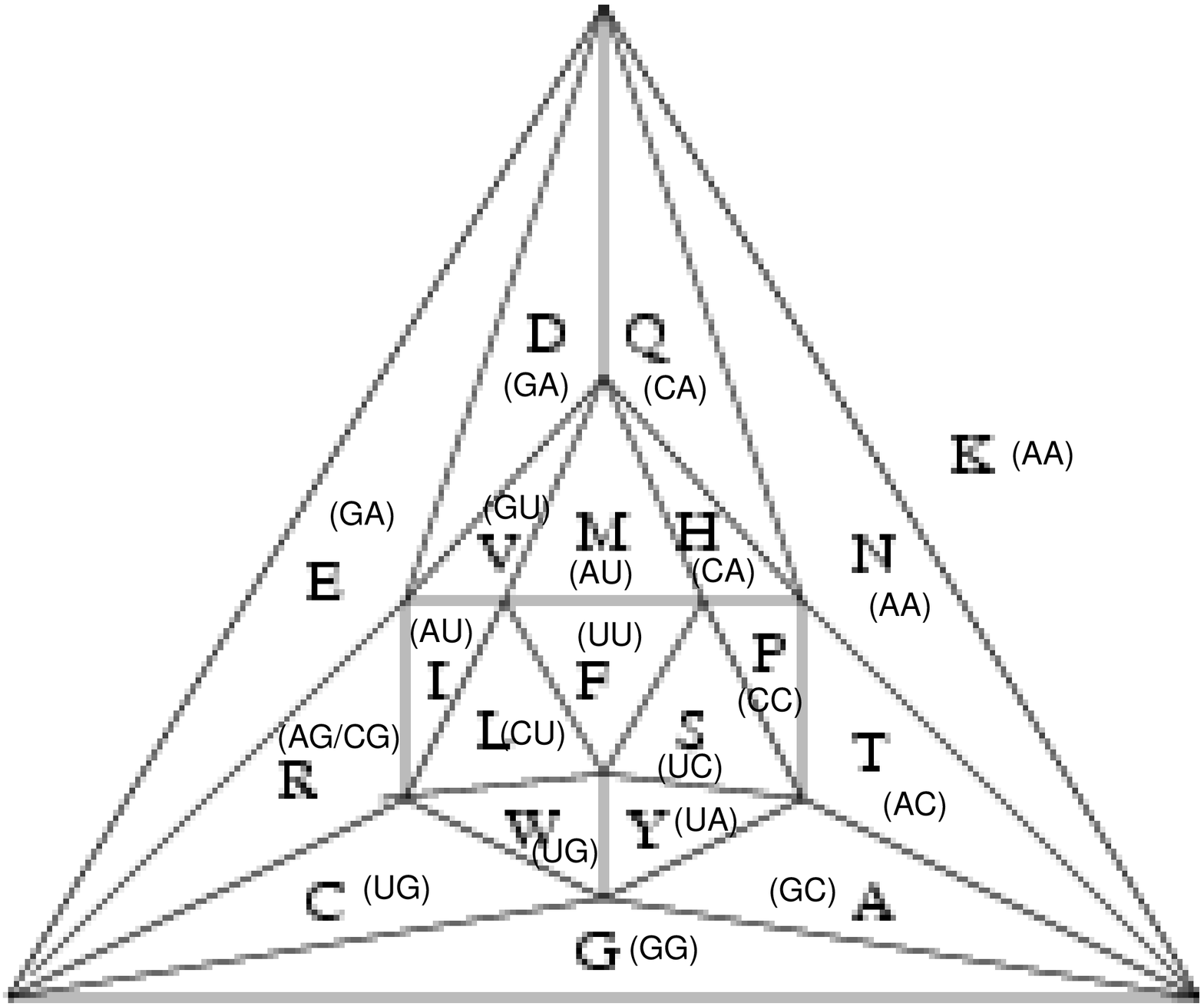}}
\subfigure[3D
Representation]{\label{fig:5.1-b}\includegraphics[scale=0.75]{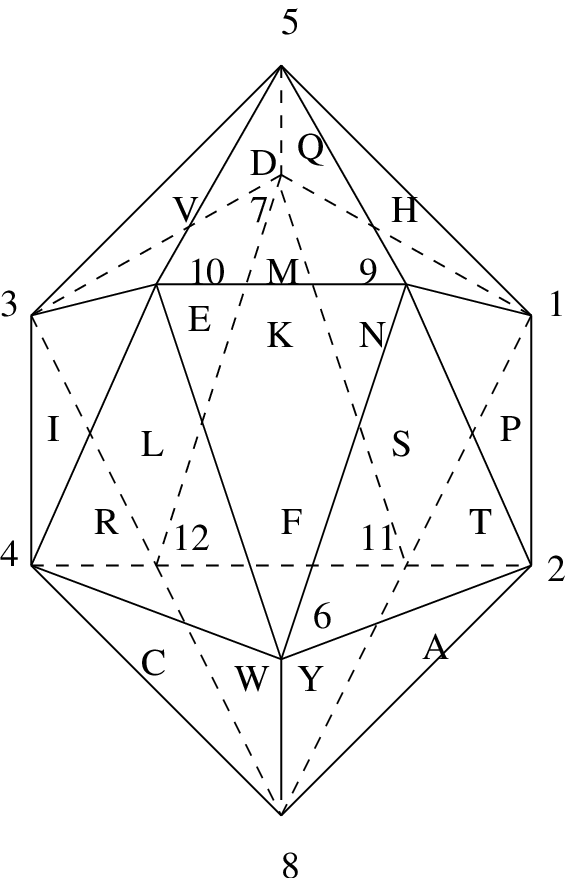}}
\end{center}
 \caption{a) 2D and  b) 3D Representation of an icosahedron with first and second nucleotide position of codon  and its amino acid mapping respectively \label{fig:5.1} }
\end{figure}
\subsection{Mapping}
\label{map}
\paragraph
\indent The amino acids are mapped onto the faces of an
icosahedron in an order based on the dinucleotide relatedness of
codons (see Chapter \ref{chap:Bioint}, section
 ~\ref{sect:gencode}). The amino acids that differ by a single
nucleotide in the codons are mapped onto the faces of an
icosahedron that are closer in three dimensional space and the
rest of the amino acids were mapped onto the faces that are
further apart. The mapping of amino acids onto two dimensional and
three dimensional representations of an icosahedron is shown in
figure 5.1a and 5.1b.
\section{Chaos game on an icosahedron}
\paragraph \indent The chaos game was played by taking the center of the
icosahedron as the starting point. When the  first amino acid is
read from a protein sequence, a point is plotted halfway between
the center of the icosahedron and the center of the face of the
corresponding amino acid. Subsequent points are plotted halfway between
the previous point plotted and the center of the
face of the amino acid read.
\paragraph
\indent Figure 5.2a shows the chaos game of a sample sequence
'MSDEFGHR' plotted. The first point is plotted halfway between the
center of the icosahedron and the center of the face corresponding
to the amino acid 'M', the second point is plotted halfway between
the first point plotted and the center of the face corresponding
to the amino acid 'S', the third point is plotted halfway between
the second point and the center of the face corresponding to the
amino acid 'D'; similarly E, F, G, H and R are plotted.
\begin{figure}[!htp]
\begin{center}
\subfigure[Chaos game on icosahedron for protein sequence
'MSDEFGHR']{\label{fig:5.2-a}\includegraphics[width=4in,height=3in]{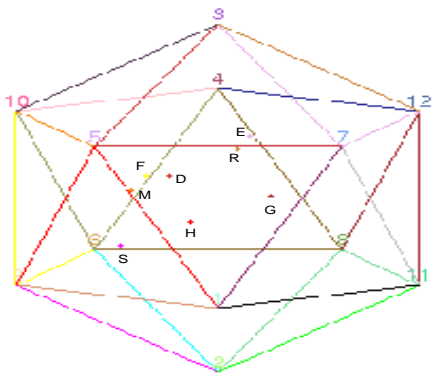}}
\subfigure[Chaos game on icosahedron for protein sequence - DNA
Polymerase Human Alpha
chain]{\label{fig:5.2-b}\includegraphics[width=4in,height=3in]{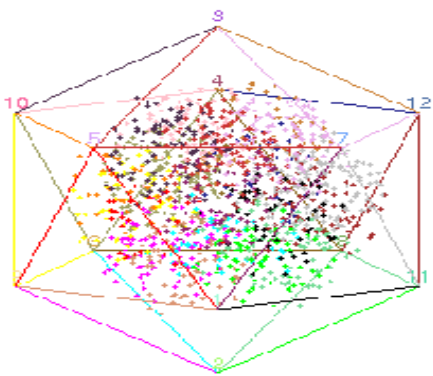}}
\end{center} \caption{ Chaos game of protein sequence in three dimension} \label{fig:5.2}
\end{figure}
\paragraph
\indent The chaos game on a protein sequence produces a
\textit{cloud of points} in space. The \textit{cloud of points}
did not reveal any obvious patterns to the naked eye. This could
be due to the length of protein sequences (approx. $< 2000$) and
also due the points being in 3D space. Figure 5.2b is the output
of chaos game using an icosahedron when played on the protein
sequence of DNA Polymerase human alpha chain (DPOA$\_$HUMAN) of length 1462. Twenty unique colors have been used to represent the twenty amino acids.
\paragraph
\indent 3D-CGR was expected to reveal useful patterns in protein sequences, since CGR has been a visual representation technique to study sequence similarities.
But, due to the points being in space no patterns were visible to the naked eye. Therefore, we reduced the number of amino acids by grouping them based on
conservative substitutions. The groupings were then mapped onto the 12 vertices of the icosahedron and chaos game was played. The points generated by the
chaos game showed empty regions near certain groups of aminoacids and more points towards other groups of amino acids indicating the frequency of occurences of the
amino acid groups but, failed to reveal any visible patterns to the naked eye. Also, we tried to visually study the points for patterns by rotating the three dimensional
figure and projecting the points onto their corresponding faces but, they were not helpful in revealing any visible patterns. Therefore, we decided to quantitatively analyse
the similarities and differences between the points produced by different sequences using phylogenetic trees.
\section{Distance measure} \paragraph \indent In this section we quantitatively analyse the relationship
between protein sequences  as well as the membership of a given protein to a protein family
by calculating the distance between the clouds of points
generated by protein sequences using 3D-CGR. In order to
define a distance measure between the clouds of points produced by any
two sequences, the points produced by the sequences were enclosed
inside a cube. The cube was then subdivided into $n \times n\times
n$ small cubes and the density of points in each of the small
cubes was calculated. Let $P$ and $Q$ be any two sequences, and $s$ and
$t$ be length of the sequences. The density of points in each of the $n
\times n\times n$ cubes of the sequences $P$ and $Q$ are represented in matrices\\
\\Let $A_{n \times n \times n}$ and $B_{n \times n \times n}$ be the 3 dimensional matrices
 \begin{center} Density $ D_A$ of $\left(a_{ijk}\right)$ = (no. of points that fall into the cube $a_{ijk}) \times  1 / s$ \\
 Density  $D_B$ of $\left(b_{ijk}\right)$ = (no. of points that fall into the cube $b_{ijk}) \times  1 / t$ \end{center}
Dividing by the length of the sequence is to normalize as the
protein sequences compared are of various length. The \textit{image distance} \cite{Wang2004} between two sequences $P$ and $Q$ is defined as\\
  $$\sum_{i,j,k=1}^n  \left| D_A \left(a_{ijk}\right ) - D_B \left(b_{ijk}\right)\right |$$
Two other distances used for sequence comparison in this thesis are the \textit{Euclid and Pearson distance}. The Euclid and Pearson distance between sequences $P$ and $Q$  using density matrices $A$ and $B$ is given by
\\The Euclid distance
  $$ \sqrt{\sum_{i,j,k=1}^n \left(D_A \left(a_{ijk}\right ) -
  D_B  \left(b_{ijk}\right)\right)^{2}}$$
\\The Pearson distance
$$ \frac{\sum_{i,j,k=1}^n D_A \left(a_{ijk}\right ) \times D_B \left(b_{ijk}\right) - \frac{\sum_{i,j,k=1}^n D_A \left(a_{ijk}\right ) \times \sum_{i,j,k=1}^n D_B \left(b_{ijk}\right )}{\mathcal{N}}}{\sqrt{(\sum_{i,j,k=1}^n D_A \left(a_{ijk}\right )^2 - \frac{(\sum_{i,j,k=1}^n D_A \left(a_{ijk}\right ))^2}{\mathcal{N}}) \times (\sum_{i,j,k=1}^n D_B \left(b_{ijk}\right )^2 - \frac{(\sum_{i,j,k=1}^n D_B \left(b_{ijk}\right
))^2}{\mathcal{N}})}}$$
$$\mathcal{N} = n \times n \times n$$
\section{Experimental objectives}
\paragraph \indent To detect protein homology using the 3D-CGR approach, to assess the
impact of dinucleotide bias at amino acid sequence level on 3D-CGR
derived protein homology and to detect shuffled  motifs, the
following experiments were performed :
\begin{itemize}
    \item \textbf{validate tree:} The goal of this experiment was to test if the phylogenetic tree constructed using 3D-CGR can detect protein sequence homology.
    \item \textbf{effect of mapping change: } The goal of this experiment was to investigate whether or not varying the mapping of the 20 amino acids on to the 20 faces of the icosahedron has an effect on the quality of the phylogenetic trees.
    \item \textbf{compare trees: } The goal of this experiment was to compare the phylogenetic trees generated by 3D-CGR with an alignment technique CLUSTALW used for studying sequence relatedness.
    \item \textbf{compare distance measures: } The goal of this experiment was to compare the phylogenetic trees generated by three distance measures and decide which one gives better results in describing the protein sequence homology.
    \item \textbf{assess fractal pattern: } The goal of this experiment was to compare the fractal patterns produced by protein sequences using their fractal dimension.
\end{itemize}
\section{ Dataset for protein sequence analysis using 3D-CGR}
\paragraph \indent The test data  was obtained from the SWISS-PROT Database. Table
5.1 and 5.2 lists all the protein sequences, their length and
SWISS-PROT ID used for the test analysis. The protein sequences
were selected such that they were of various lengths, and from
protein families of diverse functionalities.
\begin{table}[!htp]
\begin{center}
\begin{tabular}{|l|l|l|}
\hline
Protein Family & $SWISS\_PROT ID$ & Length\\
\hline
Myoglobin & $MYG\_ALLMI$(Alligator)   &   154\\
&$MYG\_CHICK$   &   153\\
&$MYG\_HUMAN$   &   153\\
&$MYG\_MOUSE$   &   153\\
\hline
Hemoglobin & $HBA\_ALLMI$   &   141\\
&$HBA\_CHICK$   &   141\\
&$HBA\_HUMAN$   &   141\\
&$HBA\_MOUSE$   &   141\\
&$HBA\_XENTR$(Frog)   &   141\\
&$HBA\_BRARE$(Fish)   &   141\\
\hline
Superoxide dismutase & $SOD1\_ORYSA$(Rice)  &   151\\
&$SODC\_DROME$(Fruit Fly)  &   152\\
&$SODC\_CHICK$  &   153\\
&$SODC\_NEUCR$(Fungus)  &   153\\
&$SODC\_XENLA$  &   150\\
&$SODC\_BRARE$  &   154\\
&$SODC\_HUMAN$  &   153\\
&$SODC\_MOUSE$  &   153\\
&$SODC\_CAEEL$(Worm)  &   158\\
&$SODC\_YEAST$  &   153\\
\hline
Alcohol dehydrogenase & $ADH1\_YEAST$  &   347\\
&$ADH1\_NEUCR$  &   353\\
&$ADH1\_BACST$  &   337\\
&$ADH1\_CAEEL$  &   349\\
&$ADH1\_ORYSA$  &   376\\
&$ADHA\_HUMAN$  &   374\\
&$ADHA\_MOUSE$  &   374\\
&$ADHA\_CHICK$  &   375\\
&$ADH1\_ALLMI$  &   374\\
\hline
\end{tabular}
\caption{Protein Family, Swiss-Prot ID and Length of test protein
sequences} \label{Table: 5.1}
\end{center}
\end{table}
\begin{table}[!htp]
\begin{center}
\begin{tabular}{|l|l|l|}
\hline
Protein Family & $SWISS\_PROT ID$ & Length\\
\hline
Catalase & $CAT1\_CAEEL$  &   524\\
&$CATA\_BACSU$(Bacteria)  &   482\\
&$CATA\_DROME$  &   506\\
&$CATA\_HUMAN$  &   526\\
&$CATA\_MOUSE$  &   526\\
&$CATA\_BRARE$  &   526\\
&$CATA\_ORYSA$  &   491\\
&$CAT1\_NEUCR$  &   736\\
&$CATA\_YEAST$  &   515\\
\hline
Methionine adenosyltransferase & $METK\_BACSU$  &   400\\
&$METK\_CAEEL$  &   404\\
&$METK\_DROME$  &   408\\
&$METK\_HUMAN$  &   395\\
&$METK\_RAT$    &   395\\
&$METK\_ORYSA$  &   396\\
&$METK\_NEUCR$  &   395\\
&$METK\_YEAST$  &   381\\
\hline
6 phosphogluconate dehydrogenase & $6PGD\_BACSU$  &   468\\
&$6PGD\_CANAL$(Yeast)  &   517\\
&$6PG1\_YEAST$  &   489\\
&$6PGD\_DROME$  &   481\\
&$6PGD\_HUMAN$  &   482\\
&$6PGD\_MOUSE$  &   482\\
\hline
DNA Polymerase &$DPO1\_BACST$  &   876\\
&$DPOA\_DROME$  &   1488\\
&$DPOA\_HUMAN$  &   1462\\
&$DPOA\_MOUSE$  &   1465\\
&$DPOA\_YEAST$  &   1468\\
&$DPOA\_ORYSA$  &   1243\\
&$DPOD\_CAEEL$  &   1081\\
\hline
\end{tabular}
\caption{Protein Family, Swiss-Prot ID and Length of test protein
sequences (continued from Table 5.1)} \label{Table: 5.2}
\end{center}
\end{table}
\section{Software}
\paragraph \indent The plotting of amino acids and evaluation of distance
measures were performed using Maple 9. X Windows was used for
running Maple 9 under the Unix Operating System. Phylip 3.63
package was used for generating phylogenetic trees to determine
the sequence similarity and differences. Fractal analysis was
performed using Java. CLUSTALW for the test data was run using the
default parameter, gap open penalty of 10 and the BLOSUM matrix.
\section {Results and discussion}
\subsection{Tree validation}
\paragraph \indent This experiment was performed to test whether the three
dimensional CGR could generate a phylogenetic tree that can
identify relatedness of sequences and distinguish differences
between sequences. The method was to use a set of sequences for
which the true phylogenetic tree was known and compare the tree
obtained by using 3D-CGR with the true phylogenetic tree. As in
nature the true phylogenetic tree is never known for sure, we used
a starting sequence and simulated its evolution through mutations
such that the relatedness of the subsequent sequences was
completely transparent.In order to perform this experiment 15
simulated protein sequences of length 352 with known percentage of
relatedness between them were created. Sequence 1 was assumed to
be the root, sequences 2 and 3 were derived from sequence 1 by 16
amino acid substitutions (in the first half for sequence 2 and
second half for sequence 3), sequences 4 and 6 were derived from
sequence 2 and sequences 5 and 7 were derived from sequence 3 by
the same method, and similarly, sequences 8 and 10 from sequence
4, sequences 12 and 14 from sequence 6, sequences 9 and 11 from
sequence 5 and sequences 13 and 15 from sequence 7. At each stage
of derivation additional substitutions were made to the derived
sequences as in the first step and the substitutions were made
such that they do not replace an earlier substitution. Figure 5.3a
represents the sample sequence derivation explained above and
figure 5.3b represents the true phylogenetic tree that the
simulated sequences were expected to generate. In parallel, the
chaos game was played on the icosahedron and the image distance
was calculated for all pair of sequences from the cloud of points
generated by their CGR's. The distances were represented in a
distance matrix and the phylogenetic tree was generated.
\subsubsection{Result and interpretation}
\paragraph \indent Figure 5.3c represents the phylogenetic tree generated by
3D-CGR. The phylogenetic tree generated by the simulated protein
sequences using 3D-CGR was able to establish sequence relatedness
based on the mutational difference in the sequences. The
hierarchical structure of ancestor and children was exact to that
of the known tree. The result shows that the 3D-CGR can identify
related sequences and distinguish differences between them.
Therefore, 3D-CGR as an analysis tool can be used to study protein
sequence relatedness.
\begin{figure}[!htp]
\begin{center}
\subfigure[SEQUENCE02 and SEQUENCE03 derived from first and second
half of SEQUENCE01, SEQUENCE04 and SEQUENCE06 derived from
SEQUENCE02; substitutions are represented in lowercase letters
]{\label{Fig:5.3a}\includegraphics[scale=0.48]{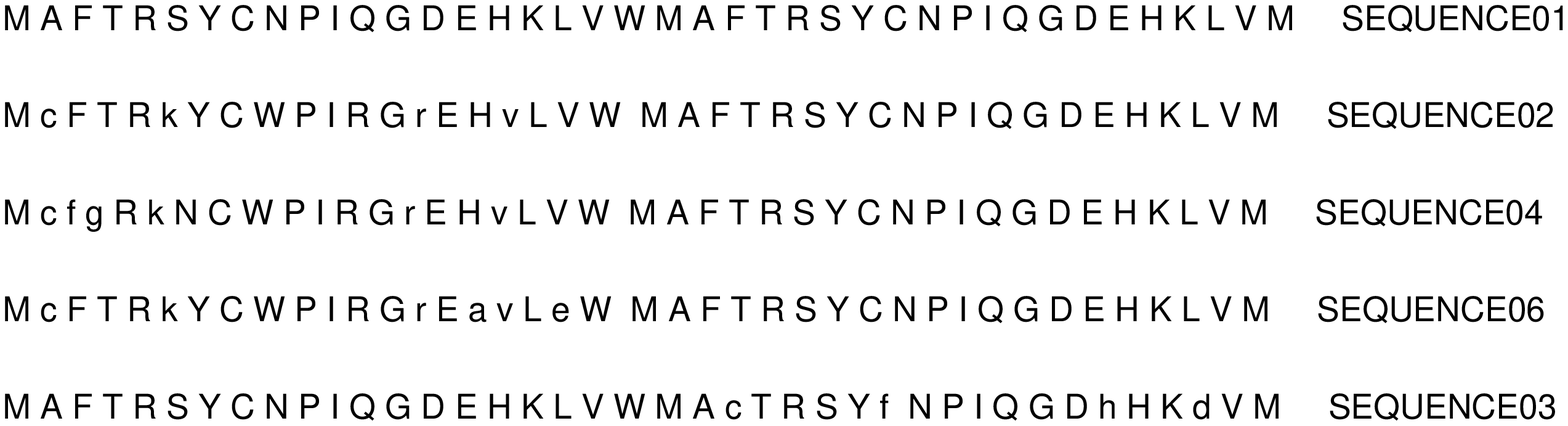}}
\subfigure[True Tree]{\label{Fig:5.3b}\includegraphics[scale
=0.70]{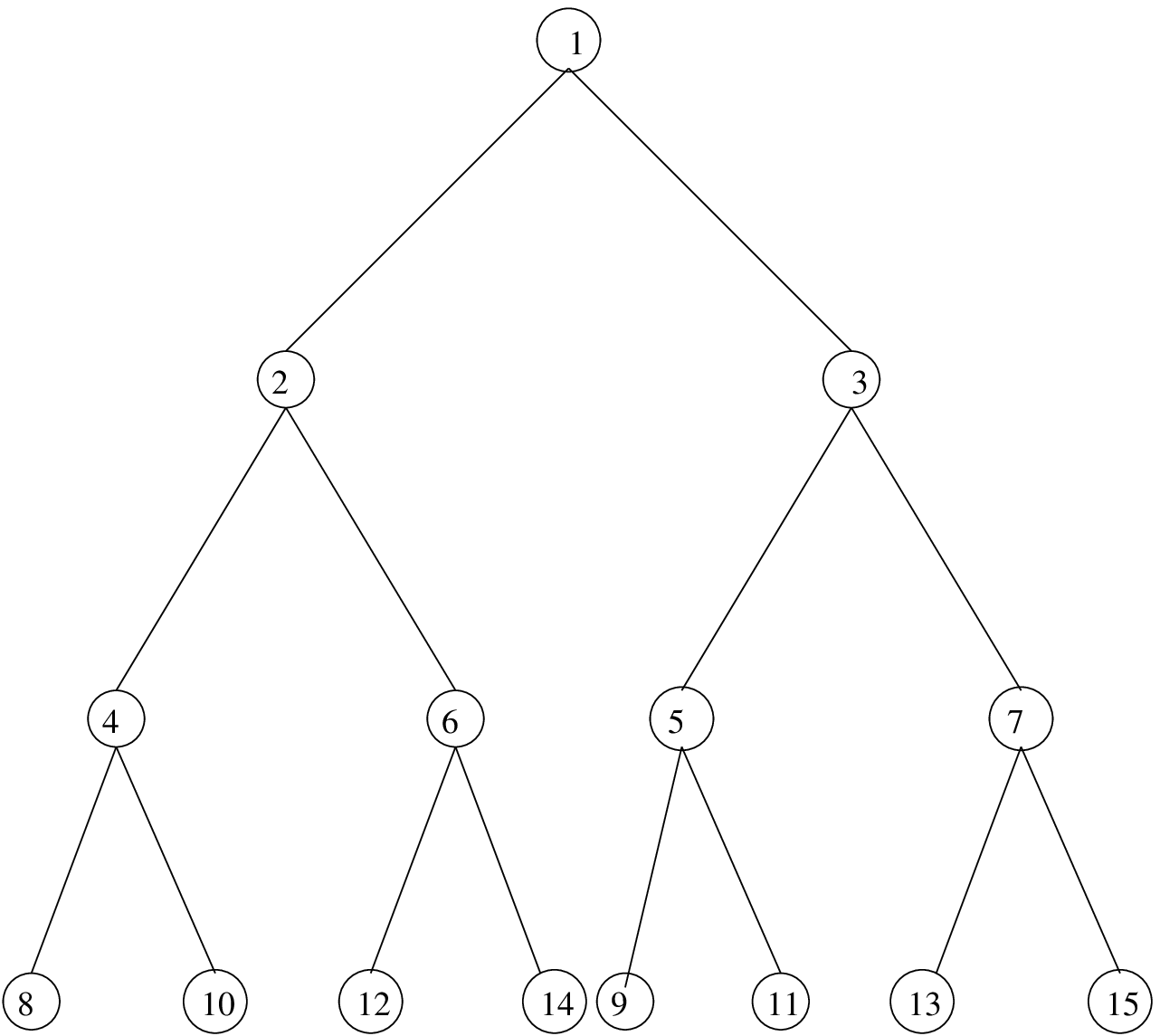}} \subfigure[Validated Tree by
3DCGR]{\label{Fig:5.3c}\includegraphics[scale=0.40]{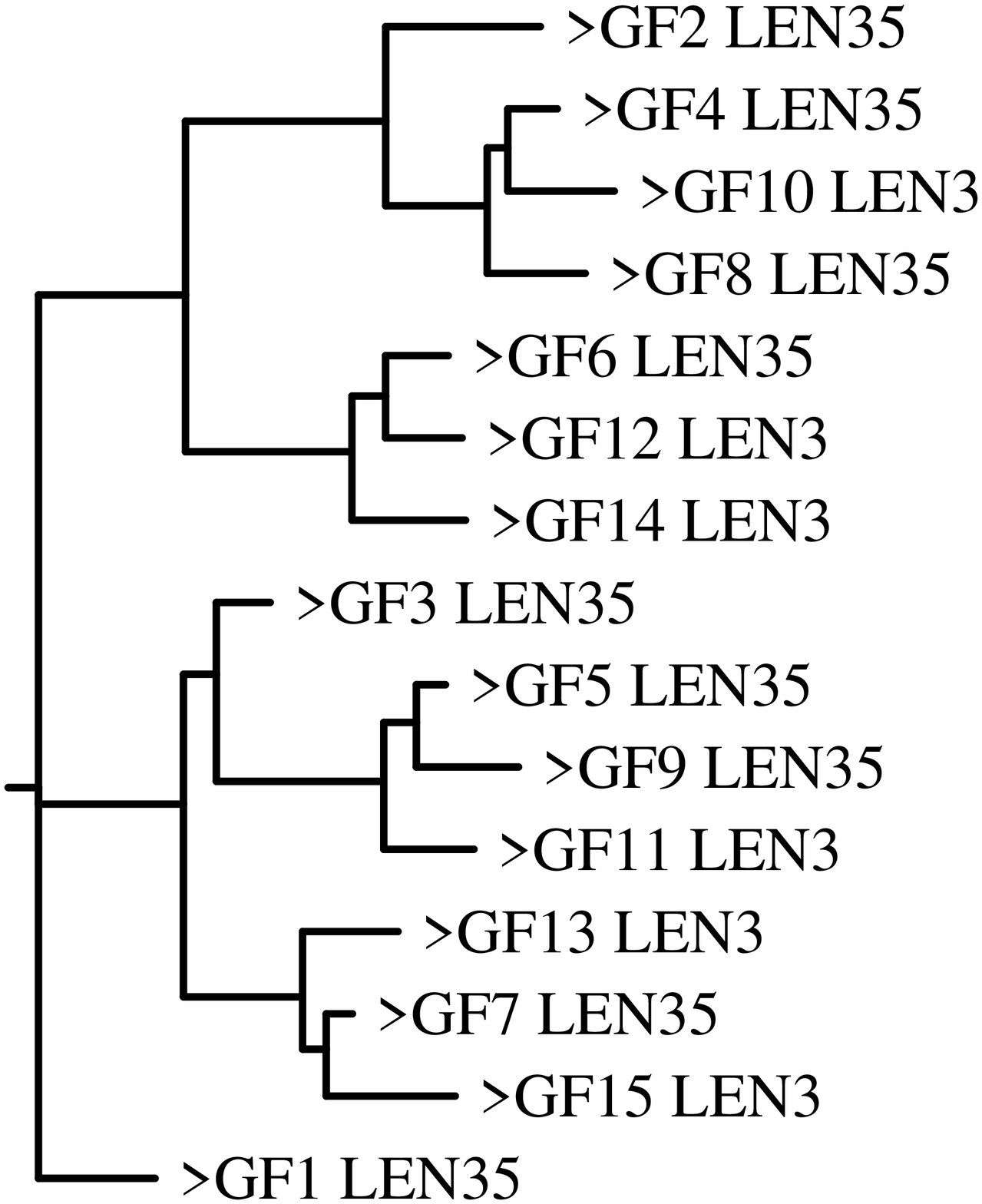}}
\end{center}
\caption{Phylogenetic tree validation -  a) Sequence derivation b)
Known tree c) Tree generated by 3D-CGR} \label{Fig:5.3}
\end{figure}
\subsection{Effect of mapping change}
\paragraph \indent In order to obtain meaningful results we wanted to map the amino acids onto the
faces of an icosahedron in a way that is biologically meaningful. Therefore, for our working mapping, we
maped the amino acids the differ by a single nucleotide in the codons onto the neighboring faces of an
icosahedron and rest of the amino acids were mapped onto the faces that are further apart.
We wanted to test whether varying the mapping would change the results of our analysis i.e we wanted
the impact of dinucleotide bias at the amino acid level. Consequently, the chaos game was played on the icosahedron
with the above mentioned mapping of amino acids and the image
distance was calculated between the cloud of points for all pairs
of sequences. The image distances between all pairs of sequences
was represented as a distance matrix and a phylogenetic tree was
generated based on the distance matrix. Similarly, phylogenetic
trees were generated for four other random mappings of amino acids
onto the faces of an icosahedron. Figures 5.4, 5.5 and 5.6
represents the phylogenetic trees obtained using the dinucleotide
related mapping and two random mappings.
\subsubsection{Results and interpretation}
\paragraph \indent All the trees generated distinguished the protein families
of the test sequences from one another. Also, the trees displayed
species relatedness within families. The comparison between the
three trees is provided in Table 5.3. The dinucleotide related  mapping differs
from random mapping 1 and random mapping 2 in the branching of ADH sequences,
in the branch order of the families and in branch length between closely related species. The difference in branch
length and branch order between the mapping based on dinucleotide
relatedness of codons and the two random mappings can be attributed to the minor
effect of dinucleotide biases at amino acid level.
\begin{figure}[!htp]
\begin{center}
\includegraphics[scale=0.75]{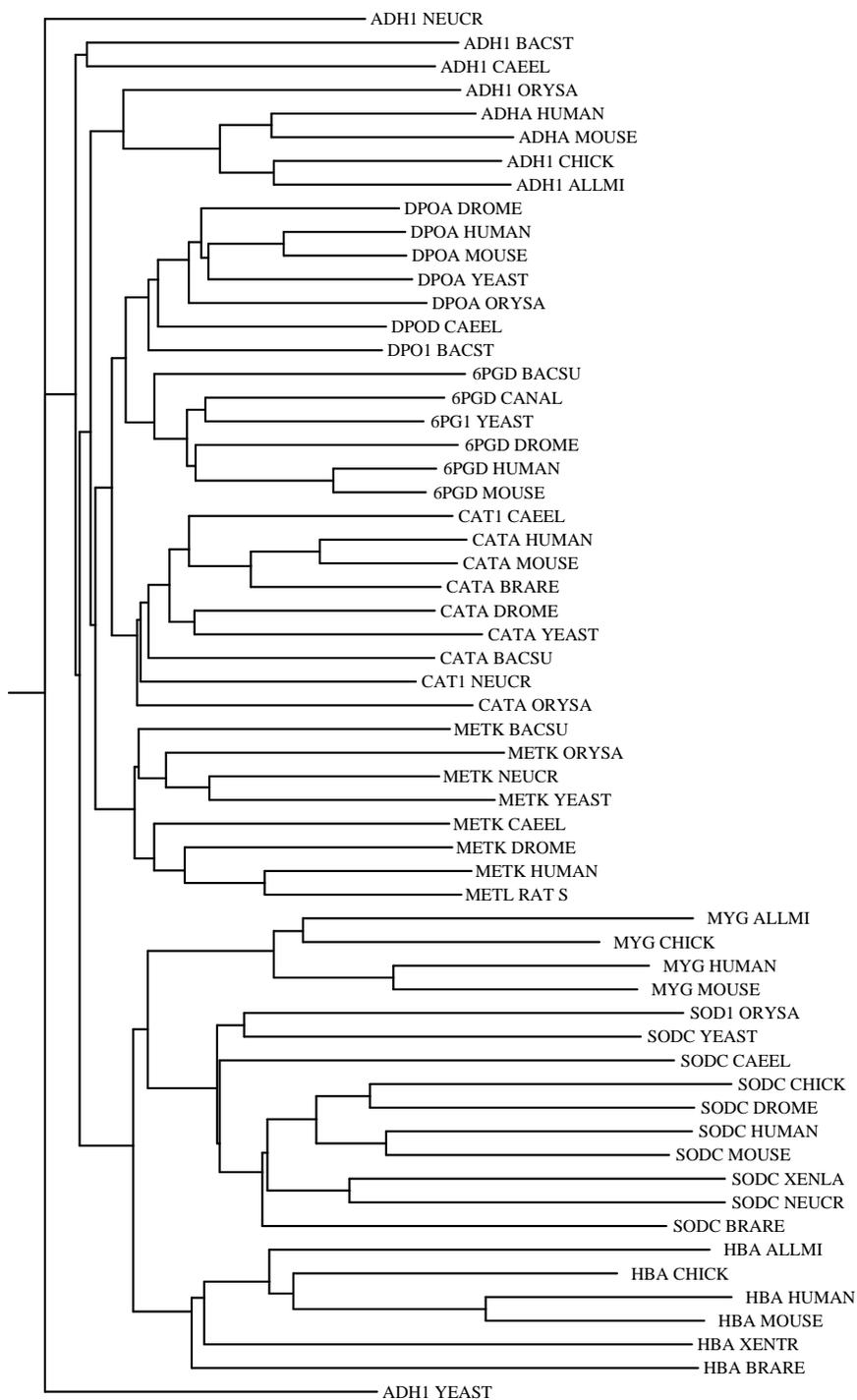}
\caption{ Phylogenetic  Tree generated by dinucleotide relatedness
mapping using the Image distance} \label{Fig:5.4}
\end{center}
\end{figure}
\begin{figure}[!htp]
\begin{center}
\includegraphics[scale=0.75]{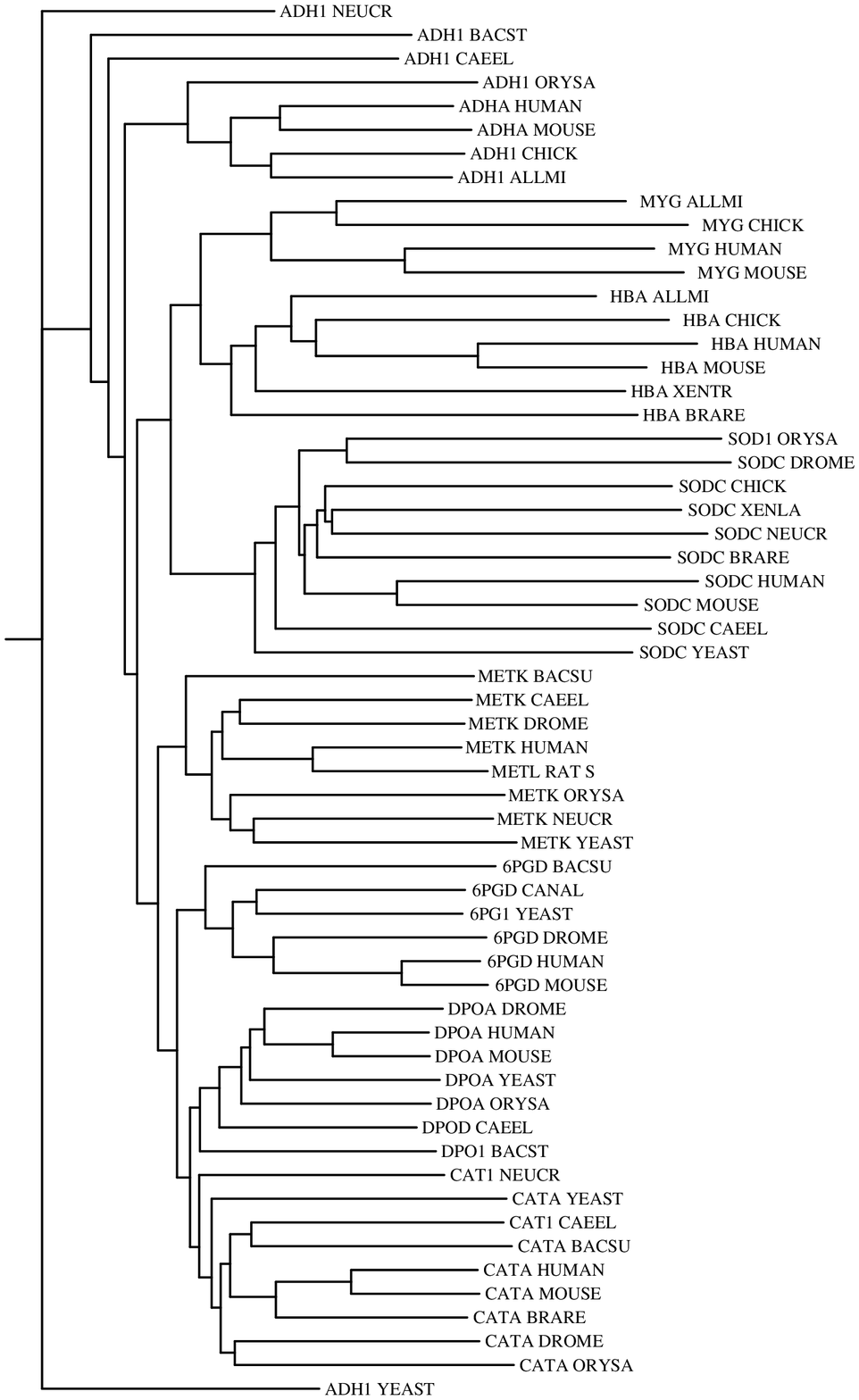}
\caption{ Phylogenetic  Tree generated by random mapping 1 using
the Image distance} \label{Fig:5.5}
\end{center}
\end{figure}
\begin{figure}[!htp]
\begin{center}
\includegraphics[scale=0.75]{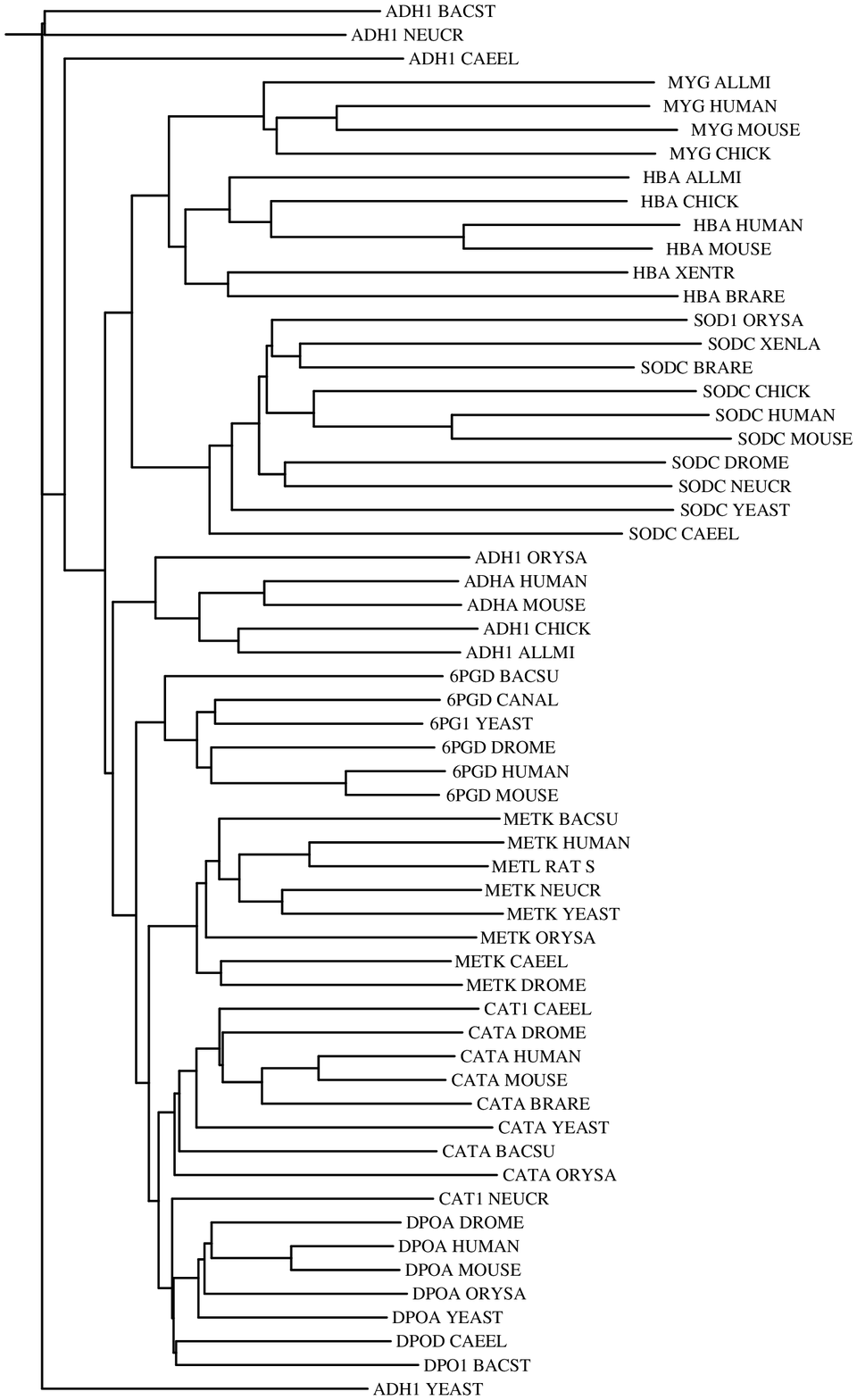}
\caption{ Phylogenetic  Tree generated by random mapping 2 using
the Image distance} \label{Fig:5.6}
\end{center}
\end{figure}
\begin{landscape}
\begin{table}
\begin{tabular}{|p{2.5in}|p{2.5in}|p{2.7in}|}
\hline
& Random Mapping 1 & Random Mapping 2\\
\hline
Dinucleotide Mapping&difference in branching of ADH sequences, difference in the evolution of protein families and difference in branch length between closely related sequences - $(ADHA\_HUMAN, AHDA\_MOUSE)$, $(MYG\_ALLMI, MYG\_CHICK)$& difference in branching of ADH sequences, difference in branch length between closely related sequences - $(ADHA\_HUMAN, AHDA\_MOUSE)$, $(MYG\_ALLMI, MYG\_CHICK)$ and evolution of protein families \\
\hline
Random Mapping 1 & identical & difference in branching of ADH sequences\\
\hline
\end{tabular}
\caption{Pairwise comparisons of the three mapping strategies} \label{Table 5.3}
\end{table}
\end{landscape}
\subsection{3D-CGR and CLUSTALW phylogenetic tree comparison}
\paragraph \indent In order to evaluate the three dimensional CGR method the phylogenetic tree generated by 3D-CGR was
compared with the tree generated by the well known multiple sequence alignment program
CLUSTALW. CLUSTALW and 3D-CGR follow different approaches to study sequence similarity.
In CLUSTALW, the similarity is initially assessed by pairwise alignment, the alignment of any pair of amino acid depends on the alignment score and is irrespective of the
pair of amino acid that preceeds it as well as the pair of amino acid that follows it. In contrast, in 3D-CGR a holistic approach is used,
every point in the image depends on the preceding sequence of points.
Phylogenetic trees were obtained for multiple sequence
alignment and 3D-CGR from the dataset. Figure 5.3 and 5.7 represent the
phylogenetic trees generated by 3D-CGR and CLUSTALW.
\begin{figure}[!htp]
\begin{center}
\includegraphics[scale=0.75]{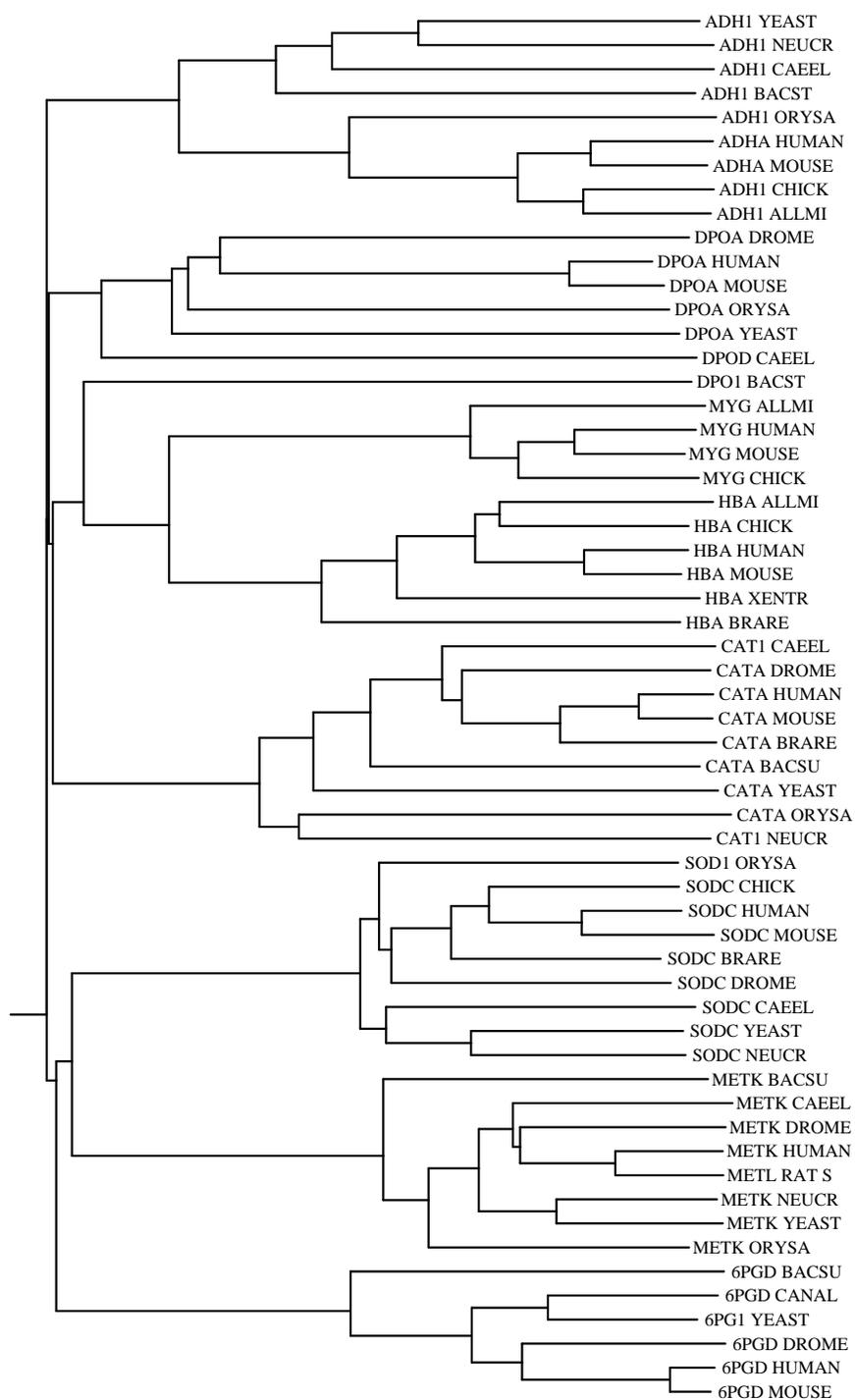}
\caption{Phylogenetic tree generated by CLUSTALW} \label{Fig:5.7}
\end{center}
\end{figure}
\subsubsection{Branch length and protein family evolution}
\paragraph \indent The trees when compared showed that they both were
able to distinguish protein families and identify species
relatedness within the familes. But the following differences were
also noted: $a$) In the CLUSTALW tree, when two sequences are
closely related (example: ADHA\_HUMAN
 and ADHA\_MOUSE), the branch lengths of the two sequences are
almost equal, whereas in the 3D-CGR tree significant differences
between the branch lengths can be seen; $b$) the evolution of
protein families in CLUSTALW tree indicate they had diverged long
time back whereas in 3D-CGR tree the evolution of protein families
indicate recent divergence and; $c$) the superoxide dismuatse
family branches off closely to myoglobin and hemoglobin in the
3D-CGR tree, whereas in CLUSTALW tree, superoxide dismutase does
not closely branch off from myoglobin and hemoglobin families.
\paragraph
\indent The branch lengths in 3D-CGR indicates 3D-CGR could be
used measure the amount of divergence of sequences within protein
families. However, we could not identify the biological
significance of the protein family evolution obtained by 3D-CGR.
\subsubsection{Shuffled Motif detection} \paragraph \indent The power of CGR relies on its holistic
approach to biological sequences. Every point on a 3D-CGR depends
on its previous point therefore it has a long memory of the
preceding amino acids in a protein sequence, whereas in sequence
alignment the alignment of any pair of amino acid does not depend
on the alignment of its preceding pair of amino acids. Therefore,
when two sequences are to be aligned and there are two different
motifs present at different positions in both the sequences (fig
5.8a), the sequence alignment would align based on the best
alignment and may not detect one of the motifs. In contrast, in
the case of 3D-CGR, the comparison of protein sequences is based on
the frequency of points in each cubic region, therefore, 3D-CGR is
expected to identify motifs better than sequence alignment. In
order to test this hypothesis, a protein sequence - SEQUENCE01 of
length 300  was selected  and a new sequence - SEQUENCE02 was
obtained from it by shuffling various regions in the SEQUENCE01.
Similarly, another sequence SEQUENCE03 was derived from SEQUENCE01
by performing several insertion/deletions and substitutions,
SEQUENCE07 was derived by interchanging two big
subsequences, and SEQUENCE06 was made to be identical to
SEQUENCE01. Fig 5.8b depicts the above example. Phylogenetic trees
for the test sequences together with two other unrelated sequences
SEQUENCE04 and SEQUENCE05 were obtained using 3D-CGR (figure 5.9a)
and CLUSTALW (figure 5.9b). In the tree obtained from CLUSTALW,
SEQUENCE03 was identified more closely to SEQUENCE01 than sequence
SEQUENCE02 whereas in 3D-CGR shuffled sequences, SEQUENCE02 and
SEQUENCE07 were identified more closely to SEQUENCE01 than
SEQUENCE02 with mutations.
\begin{figure}[!htp]
\begin{center}
\subfigure[Interchanged
subsequence]{\label{Fig:5.8a}\includegraphics[scale
=0.75]{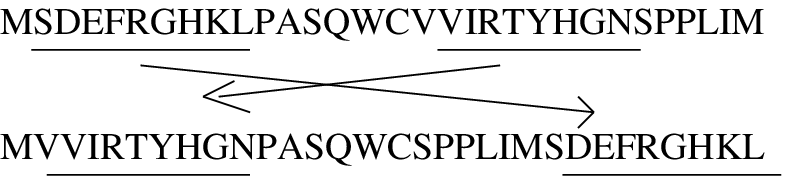}} \subfigure[Positions marked 1-8 in SEQUENCE01
are interchanged in SEQUENCE02, positions marked in SEQUENCE03 are
mutated from SEQUENCE01, SEQUENCE04 with two big interchanged
regions
marked]{\label{Fig:5.8b}\includegraphics[scale=0.50]{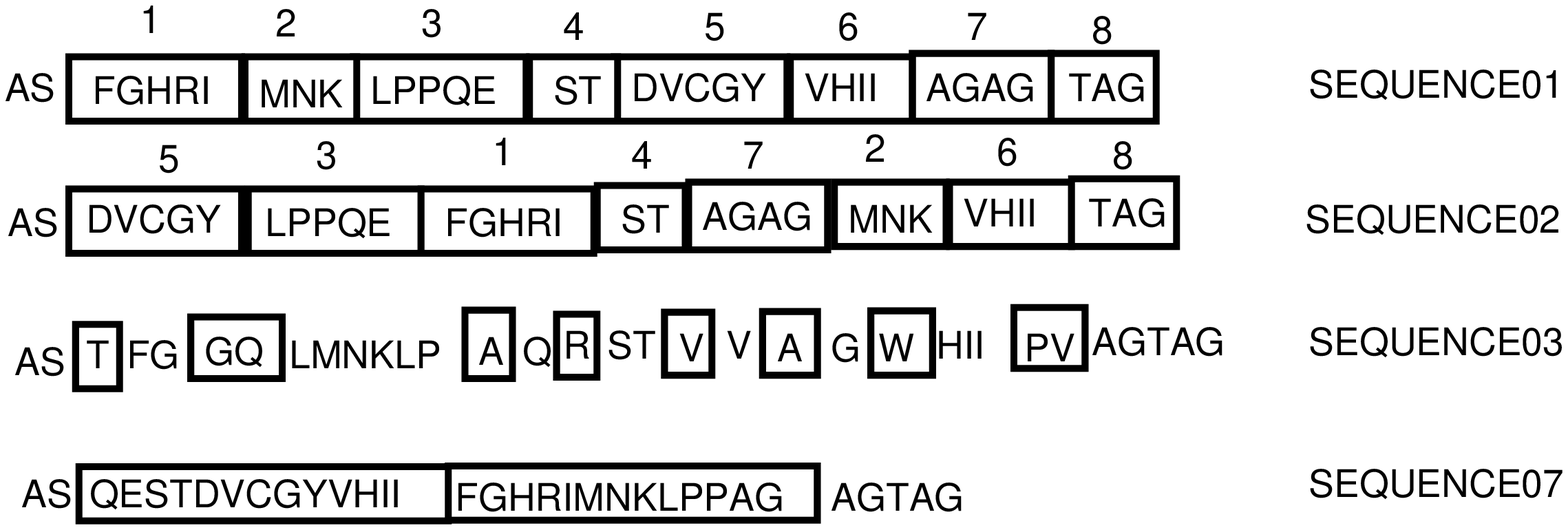}}
\end{center}
\caption{Shuffled motif detection}\label{Fig:5.8}
\end{figure}
\begin{figure}[!htp]
\begin{center}
\subfigure[Phylogenetic tree generated by Chaos for Jumbled
Sequences]{\label{Fig:5.9a}\includegraphics[scale=0.30]{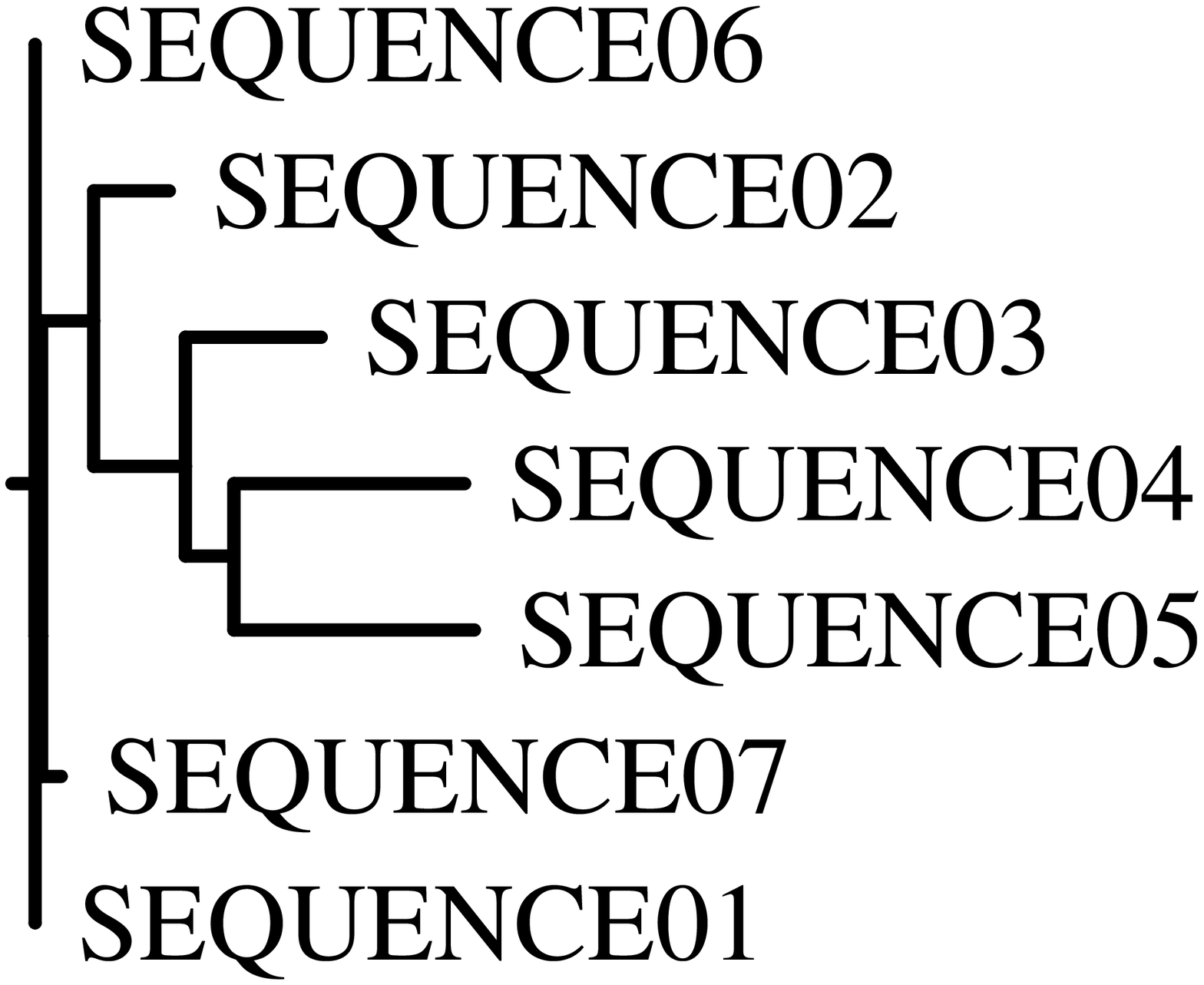}}
\subfigure[Phylogenetic tree generated by CLUSTALW for Jumbled
Sequences]{\label{Fig:5.9b}\includegraphics[scale=0.50]{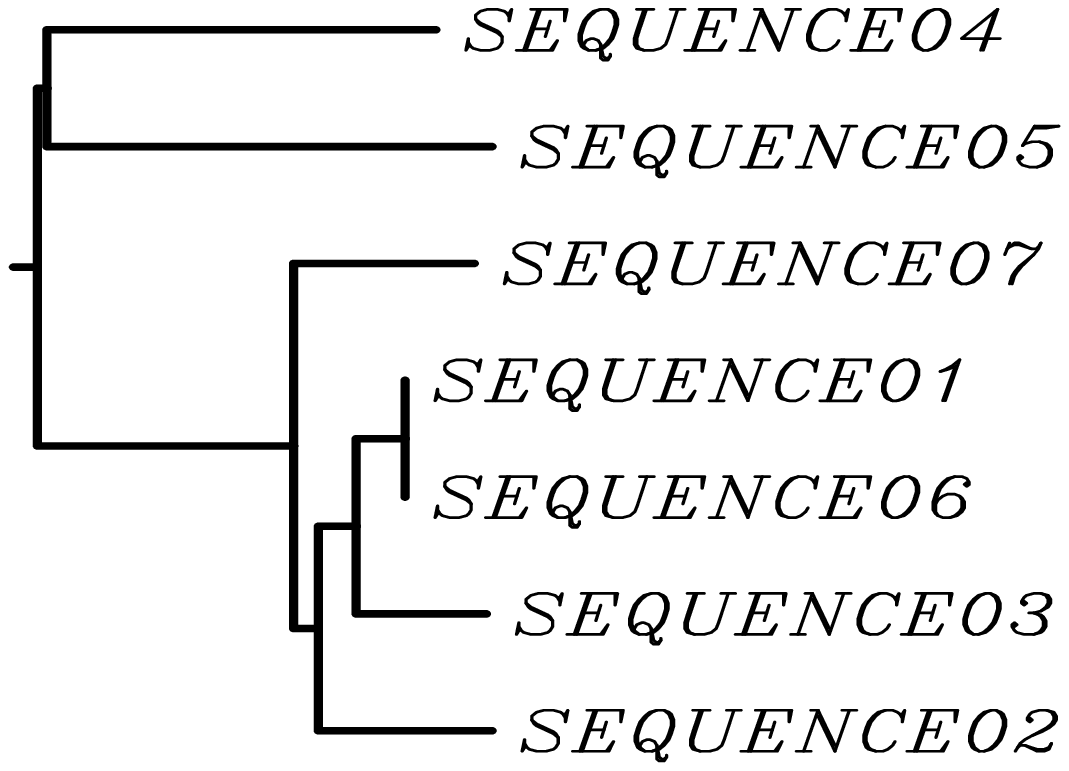}}
\end{center}
\caption{Shuffled motif detection by chaos and CLUSTALW}\label{Fig:5.9}
\end{figure}
\paragraph
\indent The above result indicates that 3D-CGR can identify shuffled motifs/domains better than CLUSTALW due to its long memory
of preceeding sequences. Therefore, 3D-CGR does not require the motifs/domains to be in the same order between sequences in contrast to CLUSTALW which requires the sequences to have same motifs/domains in the same order when assessing sequence similarity.
The ability of 3D-CGR to identify shuffled motifs/domians between any pair of sequences indicate it can be used to study protein evolution due to exon shuffling. Exon shuffling is a process by which motifs/domains have been shuffled to form new proteins.
\subsection{Distance measure comparison}
\paragraph \indent All the results obtained in section 5.7.1, 5.7.2 and 5.7.3 used the image distance.
We performed some experiments using another two different
distances in order to determine the best distance measure for the
output produced by 3D-CGR. The three distance measures used for
the comparison were the Image distance, the Euclid Distance and
the Pearson distance. The three distance matrices containing the
distances between all pairs of protein sequences were calculated
using Image, Euclid and Pearson distances and their corresponding
phylogenetic trees were generated. Figures 5.3, 5.10 and 5.11
represent the phylogenetic trees generated by the Image distance,
Euclid distance and Pearson distance measures.
\begin{figure}[!htp]
\begin{center}
\includegraphics[scale=0.75]{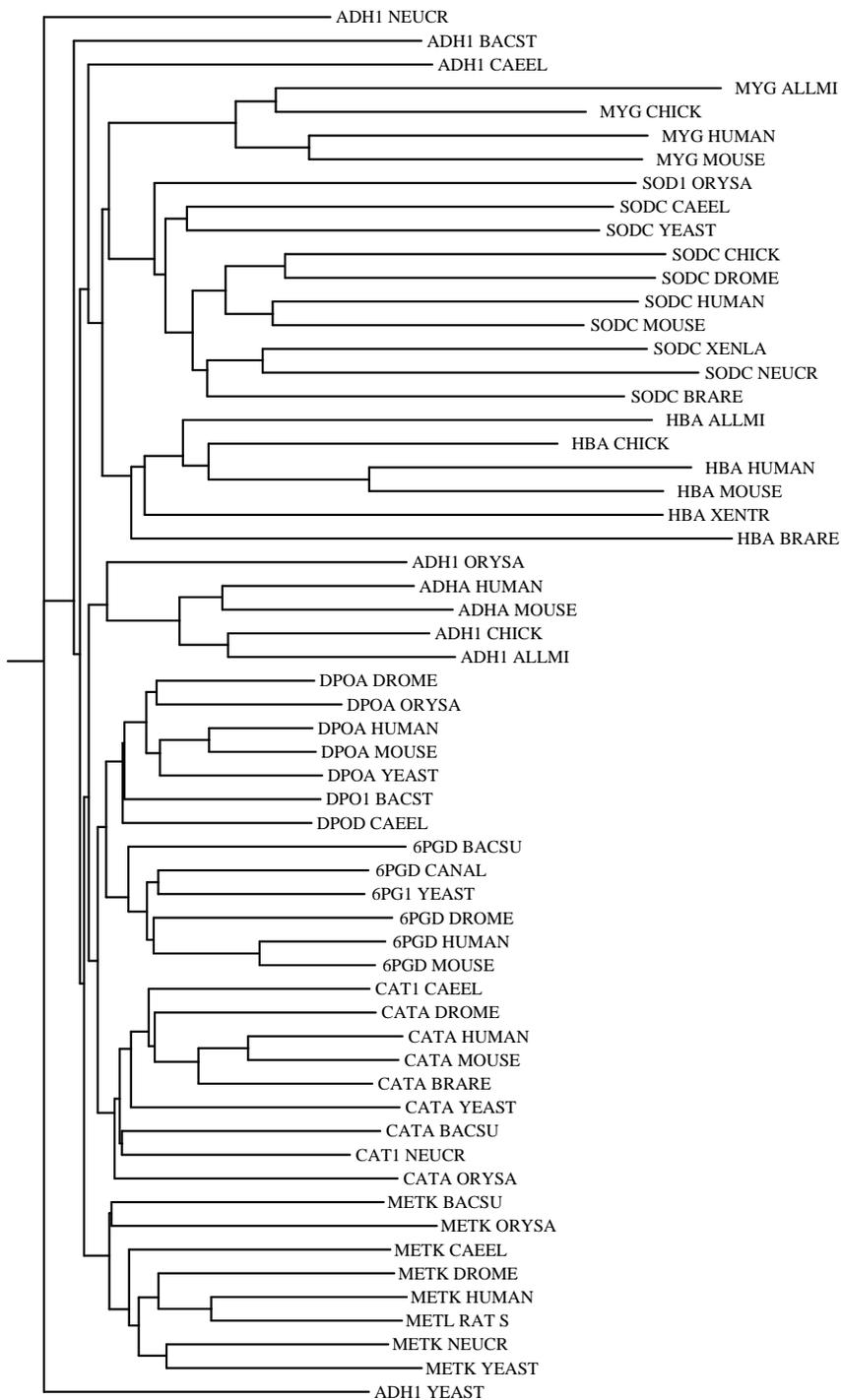}
\caption{Phylogenetic tree generated by the Euclid distance}
\label{fig:5.10}
\end{center}
\end{figure}
\begin{figure}[!htp]
\begin{center}
\includegraphics[scale=0.75]{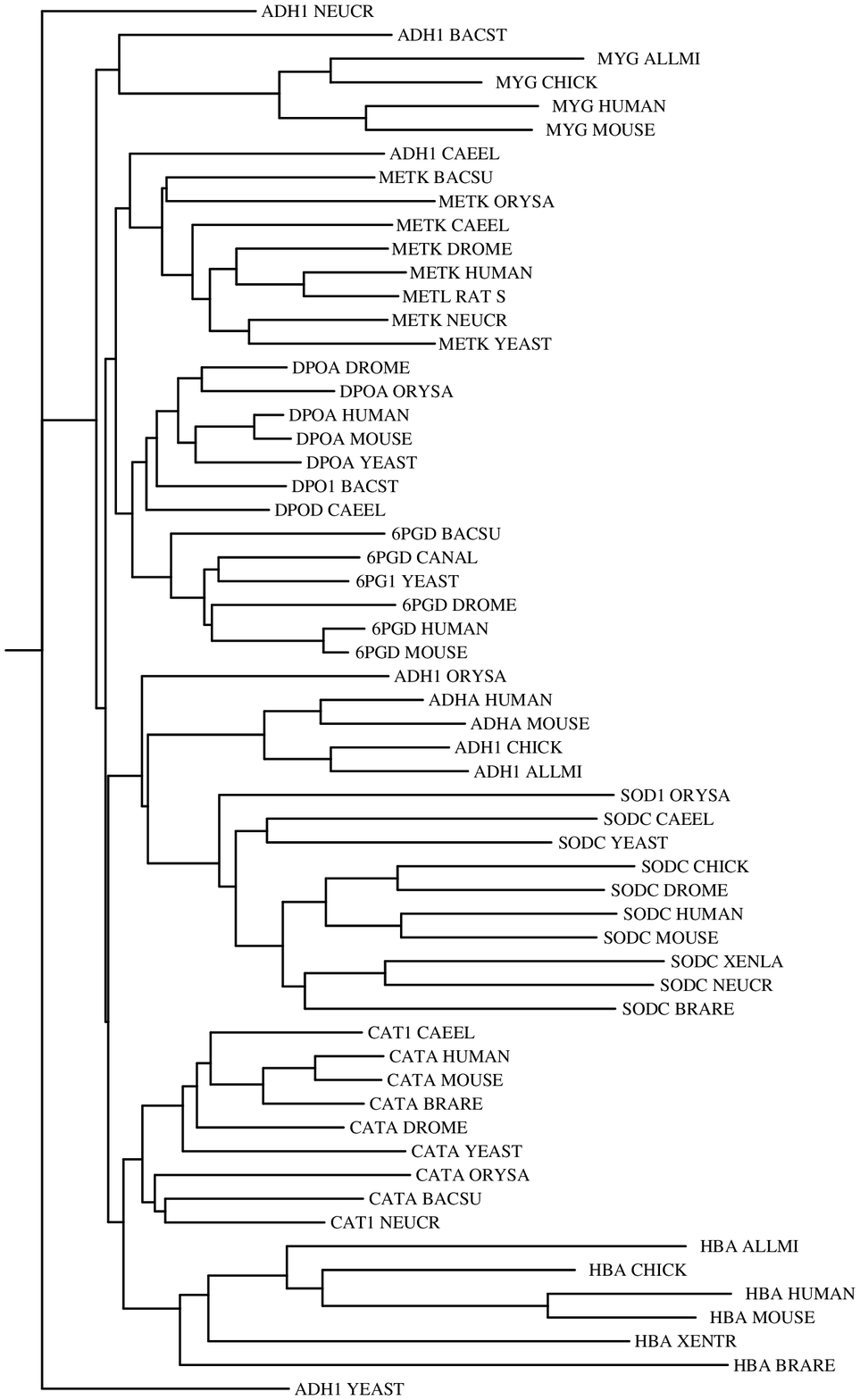}
\caption{Phylogenetic tree generated by the Pearson distance}
\label{Fig:5.11}
\end{center}
\end{figure}
\subsubsection{Result and interpretation}
\paragraph \indent The trees generated by the Image distance, Euclid distance and
Pearson Distance using 3D-CGR are all able to identify the related
species and their families. The only difference was that the tree
generated by the Pearson Distance was different from the trees generated by the Image
distance and the Euclid distance in branch order and branch length.
Therefore, it is not easy to conclude that one distance is better than
the other.
\section{Fractal analysis on protein sequences}
\paragraph \indent Fractal analysis assesses the fractal patterns produced by
a cloud of points generated by a sequence. The fractal dimension (Box Counting Dimension) is usually
used in assessing the fractal patterns. Since there was no
pattern visible to the naked eye in the cloud points produced by the
3D-CGR, the relationship between sequences was assessed by
calculating box counting dimensions for the cloud of points
generated. The fractal dimensions for various box sizes for the
sequences were calculated and the implementation was done using a
Spatial Subdivision Algorithm - Octree.
\subsection{Octree}
\paragraph
\indent
An Octree is a tree data structure with eight children used to
represent spatial subdivision. Each node of the octree holds the
physical position of the boxes. Figure 5.12 represents the
subdivided cube and its Octree. Initially, a single cube is needed
to cover the point cloud, therefore, it is the root node in the
Octree. Divide the cube into eight smaller cubes and determine if
any of the cube covers a portion of the point cloud. If they are,
then subdivide those cubes into further smaller cubes. The process
continues for some $n$ times to obtain a better approximation of the
fractal dimension.
\begin{figure}[!htp]
\begin{center}
\includegraphics[scale=0.60]{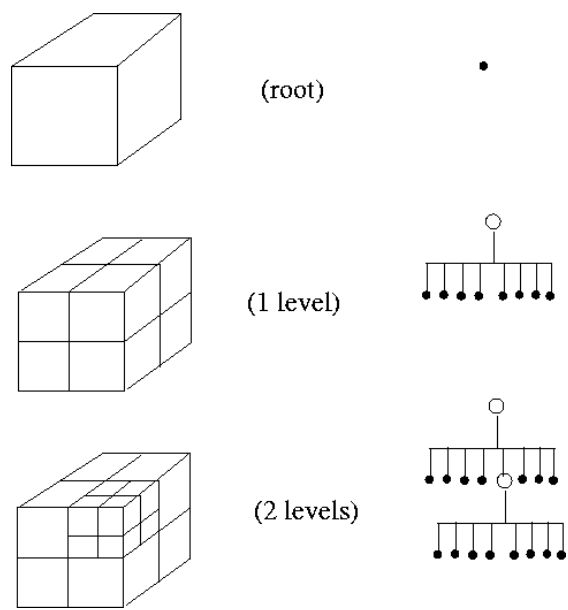}
\caption{Spatial Subdivision using Octree} \label{fig:5.12}
\end{center}
\end{figure}
\subsection{Fractal dimensions of test sequences}
\paragraph
\indent A log-log plot
of the number of boxes needed to cover the cloud of points and the box
sizes for each sequence in the data set was performed to analyze
the fractal nature of the sequences. Fig 5.13 shows the log-log
plot of the sequences in the data set.
\begin{figure}[!htp]
\begin{center}
\includegraphics[scale=0.65]{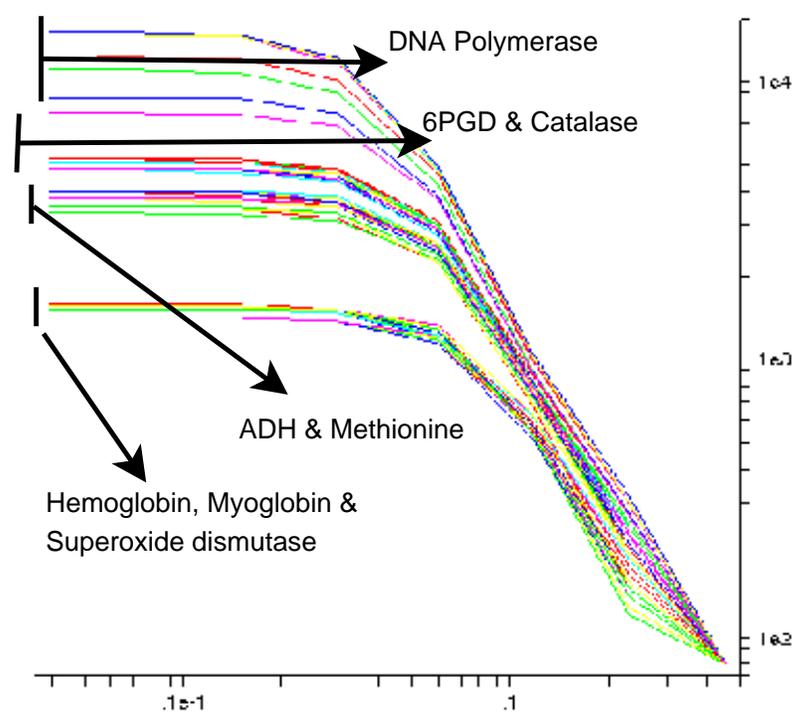}
\caption{Fractal curve for the test data set} \label{Fig:5.13}
\end{center}
\end{figure}
\subsubsection{Result and interpretation} \paragraph \indent Though the fractal
curves generated were similar for protein families, the curves were
dependent on their sequence lengths. Therefore, fractal analysis is not
good method to analyze the relationship between the
sequences.

\chapter{Conclusion}
\label{chap:conc}

\paragraph
\indent  This thesis presented the fundamentals of DNA and protein
sequences, mathematical background of chaos and fractals, two
dimensional chaos game representation while also exploring a new
approach for CGR to study sequence similarity. Chapter 3 and 4
presented the biological and mathematical background for the
thesis. Chapter 4 presented the two dimensional CGR of DNA and
protein sequences studied in the past. Chapter 5, presented the
new three dimensional approach to CGR for protein sequences that
provided a holistic approach to sequence analysis
\paragraph
\indent First, in Chapter 2, the biological introduction to the
thesis was presented. The meaning, structure and the functionality
of DNA and protein sequences and the synthesis of protein sequence
from DNA were explained. The representation of sequences/species
relatedness was explained through phylogenetic tree and the
analysis of sequences/species relatedness was explained through
the bioinformatic technique, sequence alignment. Also, the process
of multiple sequence alignment was briefly explained through a
widely used program CLUSTALW.
\paragraph
\indent Next, in Chapter 3, the mathematics background to understand the chaos game for
biological sequences was presented. The mathematical background of
chaos game for generating fractals  and its ability to reveal the
structure present in the non-random sequences were explained.
Also, the concepts of fractal and the need for fractal dimension
were explained with various examples.
\paragraph
\indent Then, in Chapter 4, the literature on chaos game
representation of DNA and protein sequences in two dimension was
presented.  A detailed description of the methods, novel advances
and limitations of chaos game on protein sequence in two dimension
was emphasized.
\paragraph
\indent Finally, in Chapter 5, a new approach and results of chaos
game representation of protein sequence in three dimension was
given. The selection of the geometric solid icosahedron to represent
twenty amino acids and its structure were explained. The new three
dimensional approach was taken to present a chaos game model by
mapping amino acids on to the icosahedron faces based on the dinucleotide
relatedness of amino acid from codons. The new approach (3D-CGR)
was used to study sequence relatedness, the effect of dinucleotide
biases at the amino acid level on the 3D-CGR deduced protein homology, and shuffled motif detection.
\paragraph
\indent The 3D-CGR was evaluated using phylogenetic trees. Trees
generated using 3D-CGR were able to distinguish protein families
and species relatedness within the families of sequences.  Also,
the effect of varying the mapping of 20 amino acids on the faces
of the icosahedron analysed using phylogenetic tree showed very
small difference in the branching of protein families and branch
lengths of closely related sequences between the phylogenetic
trees generated by random mappings and mapping based on the
dinucleotide relatedness of codon. The comparison of phylogenetic
trees of 3D-CGR and CLUSTALW revealed the significant difference
in branch length between closely related sequences indicated
3D-CGR could be used for measuring the amount of divergence
between sequences within a family. Also, 3DCGR can detect sequence
relatedness by detecting multiple motifs present in sequences
irrespective of the order of the motifs therefore, it could be a
useful tool in studying protein evolution due to exon shuffling.
\paragraph
\indent The Image distance measure used for generating
phylogenetic trees was compared with Euclid and Pearson distance.
All the three distance measures were able to distinguish protein
families and relatedness of species within the families. Finally,
sequence relatedness explored using fractal curves did not provide
much information except the fractal curves generated were similar
for protein families.
\paragraph
\indent The patterns produced by 3D-CGR were not visible to the
naked eye. Therefore, it is hoped the future research in 3D-CGR on
protein sequences could represent the 'cloud of points' in space
as a structure for visual comparison. Also, the research could be
extended to detect the position of shuffled motifs in 3D-CGR in
order to study the protein evolution due to exon shuffling.

\newpage
\begin{singlespace}
\appendix
\addcontentsline{toc}{chapter}{Appendices}
\chapter*{Appendix A}

\section*{Algorithm to determine the subsequence represented by any point in a CGR - Dutta et.al \cite{Dutta1992}}
\textit{Input}:  A point (X,Y) on the CGR.\\
\textit{Output}: Sequence that generated the point (X,Y)\\ \\
\indent Step 1. Let (0,0), (2,0),(2,2) and (0,2) be the vertices
of the square and (CX,CY) represent the center of the square. CX =
1 and CY =1\\
\\ \indent Step 2. Let (X,Y) be  the coordinates of the point whose sequence we want to determine (within the resolution limit of the monitor $\pm \delta $ of the
monitor).\\
\\  \indent  Step 3.  L is the length of the subsequence to be
generated.\\
\\  \indent Let PX and PY be two variables that hold the coordinates of the points as they are generated. Based on the center coordinate and the resolution limit of the monitor, determine in which quadrant the point belongs and change the center accordingly. The process is repeated until the X and Y values are equal to the center +/- the resolution limit of the monitor. These are given by the following steps:\\
\\  \indent  Step 4.  Repeat step 4 L times
\\  \indent \hspace{0.5cm} if  X $>$ CX and Y $> $CY then
\\ \indent \hspace{1.0cm} if X $>$ CX + $\delta$ then PX = 2
\\ \indent \hspace{1.0cm} if X $<$ CX + $\delta$ then PX = 1
\\ \indent \hspace{1.0cm} if Y $>$ CY + $\delta$ then PY = 2
\\ \indent \hspace{1.0cm} if Y $<$ CY + $\delta$ then PY = 1
\\ \indent \hspace{1.0cm} if X = CX  $\pm \delta$  or Y = CY $\pm \delta $ then goto step 6
\\ \indent \hspace{0.5cm} set CX = CX + $\left(-1\right)^{PX} $ $\left(1/2\right)$
\\ \indent \hspace{0.5cm} set CY = CY + $\left(-1\right)^{PY} $ $\left(1/2\right)$
\\ \indent \hspace{0.5cm} if PX =2 and PY = 2 then  N = G
\\ \indent \hspace{0.5cm} if PX =2 and PY = 1 then  N = T
\\ \indent \hspace{0.5cm} if PX =1 and PY = 2 then  N = C
\\ \indent \hspace{0.5cm} if PX =1 and PY = 1 then  N = A
\\ \indent Step 5. At each step set the value of CX and CY as follows, For the \textit{ith} step:
\\ \indent \hspace{0.5cm} CX = CX + $\left(-1\right)^{PX} $ $\left(1/2\right)^i$
\\ \indent \hspace{0.5cm} CY = CY + $\left(-1\right)^{PY} $ $\left(1/2\right)^i$
\\ \indent Step 6. When CX  - $\delta < X < CX + \delta$ and CY - $\delta < Y < CY + \delta$ then the length of the sequence is reached or the limit of the resolution of the CGR is
reached.\\
\\
Both CX and CY should reach the resolution limit simultaneously.
\\ The above algorithm can also be used to determine missing sequences as well.

\chapter*{Appendix B}

\section*{Algorithm to generate simulated sequence by predicting the order of nucleotides in a sequence using the probability of dinucleotide frequency - Dutta et.al in 1992  \cite{Dutta1992}}
\textit{Input}: DNA Sequence\\
\textit{Output}: CGR of simulated sequence.\\ \\
\indent  Step 1. Determine the frequency of occurrences of dinucleotides from a given
 sequence.\\
\\ \indent  Step 2. Set the length L of the hypothetical sequence to be
generated.\\
\\ \indent  Step 3. Randomly choose the first base N$\left( 1 \right)$ of the
sequence.\\
\\ \indent  Step 4. Set $i$ = N$\left( 1 \right)$, Set j = 2;\\
\\ \indent  Step 5. Repeat 6  and 7 while $j \le L$\\
\\ \indent  Step 6. Generate a number $n$ between 0 and 1.
\\ \indent \hspace{1.0cm} if 0 $<n < P_{iA}$ then N(j) = A
\\ \indent \hspace{1.0cm} if $P_{iA} < n < P_{iT} $ then N(j) = T
\\ \indent \hspace{1.0cm} if $P_{iT} <  n < P_{iG} $ then N(j) = G
\\ \indent \hspace{1.0cm} if $P_{iG} <  n < P_{iC} $ then N(j) = C
\\ \indent where  $P_{iA}$, $P_{iT}$,  $P_{iG}$ and  $P_{iC}$ represent the probability of A,T,G,C follow the $ith$ nucleotide.\\
\\ \indent  Step 7. Set i = N(j) and j = j +1.\\
\\ \indent Step 8. Generate CGR for the hypothetical sequence and compare with the CGR of the original
sequence.\\
\\ \indent  Step 9. If the CGRs do not match then reset the values of the
probabilities.\\
\\ Guidelines for setting the probabilities were  also provided based on trial and error method, also said,
the algorithm could be modified to use the probability occurrences of tri-, tetra- or $k$-nucleotides.
Therefore, \cite{Dutta1992} concluded, the sparse region in the CGRs of vertebrate gene (Fig 4.2) are due to
the rare occurrences of the dinucleotide and not due to any non-random occurrence of single nucleotide.

\end{singlespace}

%
%
%
\bibliographystyle{admplain}
\renewcommand{\bibname}{References}

\begin{singlespace}
\bibliography{samplerefs}

\begin{thebibliography}{10}
\addcontentsline{toc}{chapter}{\bibname}

\bibitem{MCW}
Bioinformatics glossary.
\newblock URL: {http://big.mcw.edu/}.\bibref{MCW}

\bibitem{PIR}
Protein information resource.
\newblock URL: {http://pir.georgetown.edu/}.\bibref{PIR}

\bibitem{Fractal}
Fractal geometry, 2005.
\newblock URL: {http://classes.yale.edu/Fractals/}.\bibref{Fractal}

\bibitem{Almedia2001}
{J.S} Almedia, {J.A} Carrico, {A.} Maretzek, {P.A.} Noble, and {M.} Fletcher.
\newblock Analysis of genomic sequences by chaos game representation.
\newblock {\em Bioinformatics}, 17(5):429--437, 2001.\bibref{Almedia2001}

\bibitem{Baranger}
{M.} Baranger.
\newblock Chaos, complexity and entropy.\bibref{Baranger}

\bibitem{Basu1997}
{S.} Basu, {A.} Pan, {C.} Dutta, and {J.} Das.
\newblock Chaos game representation of proteins.
\newblock {\em Journal of Molecular Graphics and Modelling}, 15:279--289,
  1997.\bibref{Basu1997}

\bibitem{Philip2003}
{P.} Bourne and {H.} Weissig, editors.
\newblock {\em Structural Bioinformatics}.
\newblock 2003.\bibref{Philip2003}

\bibitem{Dutta1992}
{C.} Dutta and {J.} Das.
\newblock Mathematical characterization of chaos game representation.
\newblock {\em Journal of Molecular Biology}, 228(3):715--719,
  1992.\bibref{Dutta1992}

\bibitem{Eidhammer2004}
{I.} Eidhammer, {I.} Jonassen, and {W.R.} Taylor.
\newblock {\em Protein Informatics}.
\newblock John Wiley $\&$ Sons, 2004.\bibref{Eidhammer2004}

\bibitem{Falconer1990}
{K.} Falconer.
\newblock {\em Fractal Geometry, Mathematical Foundations and Applications}.
\newblock John Wiley $\&$ Sons, 1990.\bibref{Falconer1990}

\bibitem{Fiser1994}
{A.} Fiser, {E.G.} Tusnady, and {I}. Simon.
\newblock Chaos game representation of protein structure.
\newblock {\em Journal of Molecular Graphics}, 12, 1994.\bibref{Fiser1994}

\bibitem{Goldman1993}
{N.} Goldman.
\newblock Nucleotide, dinucleotide and trinucleotide frequencies explain
  patterns observed in chaos game representations of dna sequences.
\newblock {\em Nucleic Acids Research}, 21(10):2487--2491,
  1993.\bibref{Goldman1993}

\bibitem{Harte2001}
{D.} Harte.
\newblock {\em Multifractals - Theory and Application}.
\newblock 2001.\bibref{Harte2001}

\bibitem{Hill1992}
{K.A} Hill, {N.J} Schisler, and {S.M} Singh.
\newblock Chaos game representation of coding regions of human globin genes and
  alcohol dehyrogenase genes of phylogenectically divergent species.
\newblock {\em Journal of Molecular Evolution}, 35:261--269,
  1992.\bibref{Hill1992}

\bibitem{Hill1997}
{K.A.} Hill and {S.M} Singh.
\newblock The evolution of species-type specificity in the global dna sequence
  organization of mitochondrial genomes.
\newblock {\em Genome}, 40:342--356, 1997.\bibref{Hill1997}

\bibitem{Jeffrey1990}
{J.H} Jeffrey.
\newblock Chaos game representation of gene structure.
\newblock {\em Nucleic Acids Research}, 18(8), 1990.\bibref{Jeffrey1990}

\bibitem{Krane2003}
{D.E} Krane and {M.L.} Raymer.
\newblock {\em Fundamental concepts of Bioinformatics.}
\newblock Benjamin Cummings, 2003.\bibref{Krane2003}

\bibitem{Pleibner1997}
{K.} Pleibner, {L.} Wernisch, {H.} Oswald, and {E.} Fleck.
\newblock Representation of amino acid sequences as two-dimensional point
  patterns.
\newblock {\em Electrophoresis}, 18:2709--2713, 1997.\bibref{Pleibner1997}

\bibitem{Saitou1987}
{N.} Saitou and {M.} Nei.
\newblock The neighbor-joining method: A new method for reconstructing
  phylogenetic trees.
\newblock {\em Molecular Biology Evolution}, 4(4):406--425,
  1987.\bibref{Saitou1987}

\bibitem{Wang2004}
{Y.} Wang, {L.} Kari, {K.A} Hill, and {S.M.} Singh.
\newblock The spectrum of genomic signatures: from dinucleotide to chaos game
  representation.
\newblock {\em GENE}, 346:173--185, 2005.\bibref{Wang2004}

\bibitem{Yu2004}
{Z.} Yu, {V.} Anh, and {K.} Lau.
\newblock Chaos game representation of protein sequences based on the detailed
  hp model and their multifractal and correlation analyses.
\newblock {\em Journal of Theoretical Biology}, 226:341--348,
  2004.\bibref{Yu2004}

\bibitem{Zimmermann2003}
{K.} Zimmerman.
\newblock {\em An Introduction to Protein Informatics}.
\newblock Kluwer Academic, 2003.\bibref{Zimmermann2003}

\end{thebibliography}

\newpage
\begin{center}{\huge \textbf{Vita}}\end{center}
\addcontentsline{toc}{chapter}{Vita}
\thispagestyle{plain}
\begin{tabular}{p{0.30\textwidth}p{0.60\textwidth}}
\textbf{Name} &  Annie Christiana Vasthi Thomas \\
 & \\
\textbf{Place of Birth}  &  Kancheepuram, Tamilnadu, India \\
 & \\
\textbf{Post-secondary} & Madras Christian College \\
\textbf{Education} & Tambaram, Tamilnadu, India \\
\textbf{and Degrees} & 1992--1995 B.Sc. \\
 & \\
 & Madras Christian College\\
 & Tambaram, Tamilnadu, India \\
 & 1995--1998 M.C.A. \\
 & \\
 & The University of Western Ontario \\
 & London, Ontario, Canada \\
 & 2003--2005 M.Sc. \\
 & \\
\textbf{Honours and Awards}  & Internation Student Scholarship, 2003-2004 \\
 & \\
\textbf{Related work}  &   Lecturer, Madras Christian College\\
\textbf{experience}  &  Tamilnadu , India \\
 & 2002-2003 \\
 & \\
 & Lecturer, Bharath Institute of Science and Technology \\
 & Tamilnadu, India\\
 & 2001--2002 \\
 &\\
 & Teaching Assistant \\
 & The University of Western Ontario \\
 & 2003--2004 \\
 & \\
 & Research Assistant \\
 & The University of Western Ontario \\
 & 2003--2005 \\
 & \\
\end{tabular}

\end{singlespace}
\end{document}